\input{aipcheck}

\documentclass[
    ,final            
  ]
  {aipproc}
\layoutstyle{6x9}

\begin{document}

\title{Searching for Particle Physics
Beyond the Standard Model
at the LHC and Elsewhere}


\author{John Ellis}{
  address={CERN, Theory Division, CH-1211 Geneva 23, Switzerland; \\
  King's College London, Department of Physics, Strand, London WC2R 2LS, UK}
}



\begin{abstract}
 Following a general introduction to open questions beyond the Standard Model, the prospects for
 addressing them in the new era opened up by the LHC are reviewed. Sample highlights are given
 of ways in which the LHC is already probing beyond previous experiments, including the searches
 for supersymmetry, quark and gluon substructure and microscopic black holes.

\begin{center}
CERN-PH-TH/2011-003, KCL-PH-TH/2011-05
\end{center}

\end{abstract}

\maketitle


\section{Introduction}

Particle physics is poised at the threshold of a new era. 
The Standard Model is well established, and poses a number of
well-defined questions to be addressed by forthcoming experiments.
The Large Hadron Collider (LHC) has now entered into operation with
a centre-of-mass energy of 7~TeV, and is already surpassing previous
accelerators in some of its probes of possible physics beyond the Standard Model.
In the near future, the LHC will explore new physics at the TeV scale, where the
mythical Higgs boson (or whatever replaces it) should lurk, and may also be able
to identify particles providing the astrophysical dark matter. In parallel, other
experiments complement and compete with the LHC, e.g., the Fermilab Tevatron
collider and direct searches for dark matter. One way or another, many of the
open questions beyond the Standard Model may soon be answered.

\section{Setting the Scene}

\subsection{The Standard Model rules OK}

\begin{figure}
\label{fig:SMtests}
  \includegraphics[height=.35\textheight]{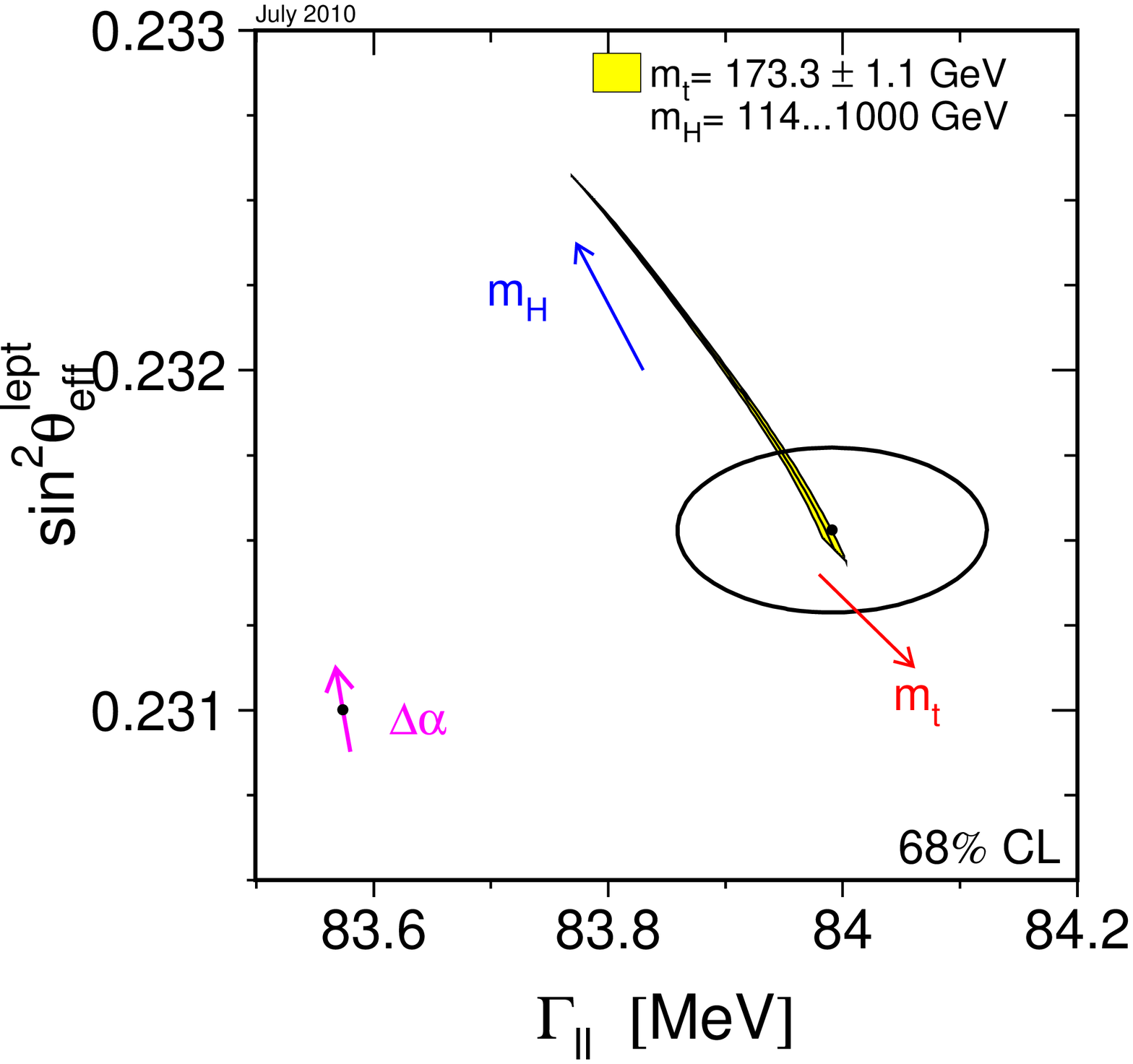}
  \includegraphics[height=.35\textheight]{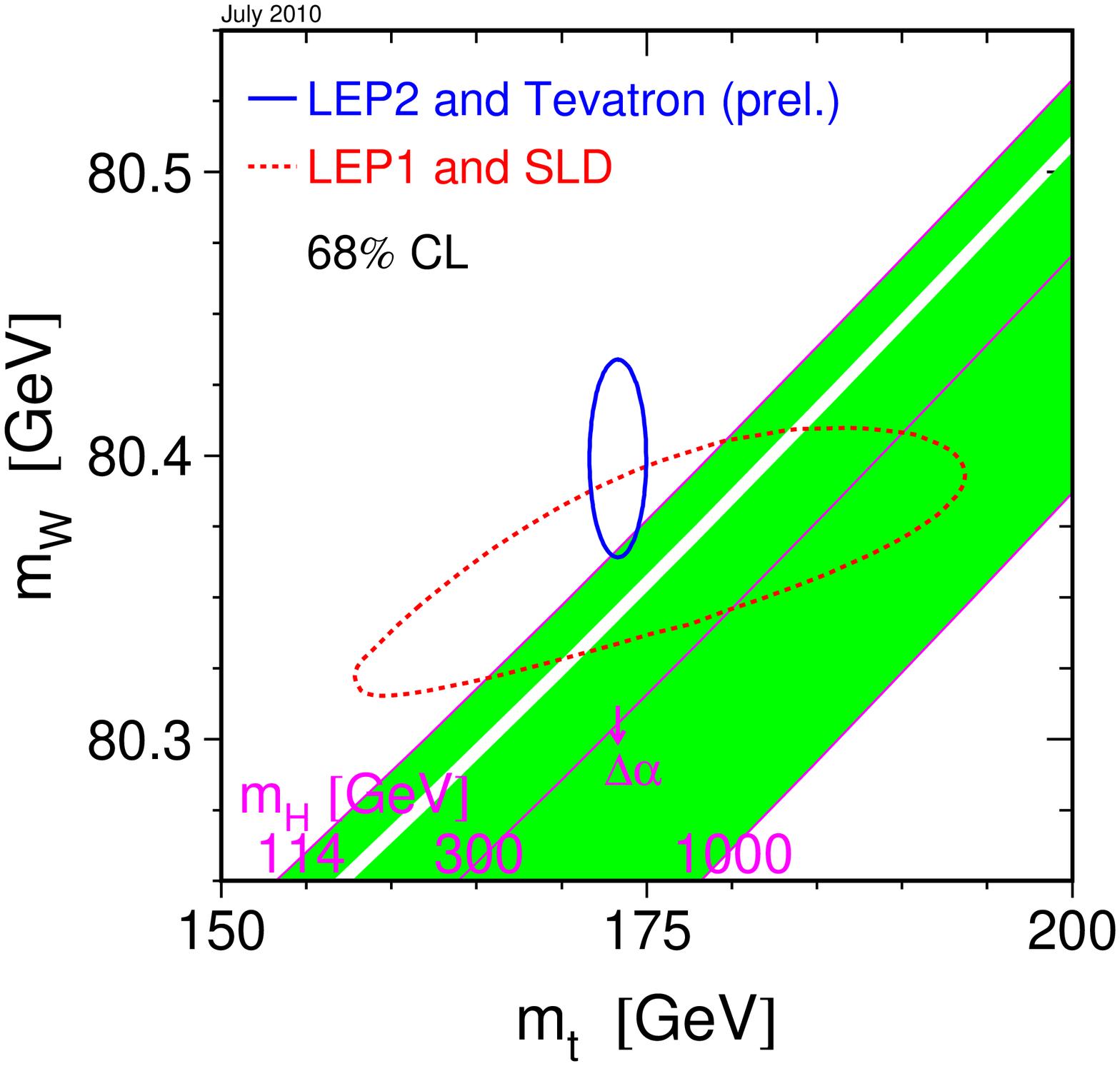}
  \caption{Precision measurements of (left) lepton couplings to the $Z^0$ and (right)
  $m_W$ and $m_t$ (solid ellipses), both of which favour a relatively low mass for the Higgs boson
  within the Standard Model~\protect\cite{LEPEWWG}. 
  In the right panel, the predictions for $m_t$ and $m_W$ based
  on low-energy measurements are shown as a mango-shaped dotted line.}
\end{figure}

The matter particles of the Standard Model~\cite{Medellin} comprise six quarks, and three charged leptons
each accompanied by its light neutrino. Four fundamental forces act on these matter
particles, namely gravity, electromagnetism and the strong and weak interactions.
With the exception of gravity, each of these forces is known to have a quantum carrier particle,
the photon, gluon, $W^\pm$ and $Z^0$ particles, respectively.
Taken together, these matter and force particles, their masses and couplings, are
sufficient to describe the results of all confirmed laboratory experiments within
their measurement accuracies. Examples of some high-precision measurements
providing checks on Standard Model predictions for the weak and electromagnetic
forces are shown in Fig.~\ref{fig:SMtests}~\cite{LEPEWWG}. Measurements of the $Z^0$ couplings to
charged leptons, shown in the left panel, agree with the Standard Model prediction at the {\it per mille} level,
and are sensitive  {\it via} quantum effects to the masses of the top quark and the hypothetical Higgs boson.
Measurements of the masses of the top quark and $W^\pm$ boson also agree very well
with Standard Model predictions based on lower-energy data, as shown in the right panel.
Interestingly, both sets of measurements seem to favour a relatively low mass
for the unseen Higgs boson, as do a number of other precision electroweak measurements.

These and many other successes inform us that the Standard Model particles can be regarded 
as the cosmic DNA, encoding the information required to assemble all the visible 
matter in the Universe.

\subsection{Questions beyond the Standard Model}

Many of the important open questions beyond the Standard Model are
already implicit in its successes~\cite{Medellin}. First and foremost may be {\it the origin of particle masses}:
are they indeed linked to a Higgs boson, or has Nature chosen a different mechanism?
As has already been mentioned, there are six quarks, three charged leptons and three
neutrinos: {\it why are there so many types of matter particles}, and why not either more or fewer?
The Standard Model describes very well the visible matter in the Universe, but
{\it what is the dark matter} in the Universe, and is it composed of elementary particles?
There are several different fundamental forces, and the electromagnetic and weak
forces are partially unified: {\it is it possible to unify} all the fundamental forces?
Finally, theoretical physicists should be deeply embarrassed that, about a century
after the discovery of quantum mechanics and general relativity, we still do not have
an established, {\it consistent quantum theory of gravity}. Maybe it could be based on string?

Each of these questions is being addressed by the LHC, which may well provide
some of the answers. For example, the search for a Standard Model Higgs boson has been
a benchmark in the design of the ATLAS and CMS detectors for the LHC~\cite{TDRs}, which
should either discover or exclude it over all the mass range up to $\sim 1$~TeV.
A dedicated experiment, LHCb, is studying CP violation and rare decays of heavy
quarks, looking for new physics beyond the dominant Cabibbo-Kobayashi-Maskawa
paradigm within the Standard Model~\cite{LHCb}. Supersymmetry and/or extra dimensions are
features of unified theories, and may also lie within the reach of the ATLAS and CMS
experiments at the LHC~\cite{TDRs}. Last but not least, detailed measurements in such theories 
might provide vital clues towards the construction of a unified quantum Theory of Everything, and
the AdS/CFT correspondence suggested by string theory may provide insights into the heavy-ion
collisions being studied by ALICE~\cite{ALICE}, ATLAS and CMS.

Of course, the LHC is not the only location for experiments addressing these
questions, and some other experimental approaches are also featured in this talk.

\section{To Higgs or not to Higgs?}

Newton taught us that weight is proportional to mass, and Einstein discovered that
energy is related to mass, but neither of these honourable gentlemen got around
to explaining the origin of mass. So where do particle masses come from?
Did Englert, Brout~\cite{EB} and Higgs~\cite{Higgs} find the answer? Are they due to the mythical Higgs 
boson, which has now become the particle physicists' Holy Grail?

\subsection{A Flaky Higgs Analogy}

For a simple analogue of the Englert-Brout-Higgs~\cite{EB,Higgs,others} mechanism and the role of the
Higgs boson, think about an infinite, flat, featureless, homogeneous and isotropic
field of snow, like the Arctic tundra in winter. Now consider trying to cross it. If you
have skis, you will not sink into (interact with) the Englert-Brout-Higgs snow field, and will 
move fast, like a particle without mass such as the photon, which always travels
at the speed of light. On the other hand, if you have snowshoes, you will sink into 
the snow (interact with the Englert-Brout-Higgs field), and move more slowly, rather
like a particle with mass such as the electron. Finally, if you have no snow equipment
apart from hiking boots, you will sink deeply into (interact strongly with) the 
Englert-Brout-Higgs snow field, like a particle with large mass such as the top quark.

So where does the Higgs boson fit into this analogy? Just as a real snow field is
made of snowflakes, and the electromagnetic field has an associated quantum
(the photon), there should be a quantum of the Englert-Brout-Higgs field, as was
first pointed out explicitly by Higgs~\cite{Higgs}. This snowflake is what we call the Higgs boson.
In the original model of Englert, Brout and Higgs the field and the quantum are 
elementary, but real life may be more complicated: just as every snowflake is a
different composite object made out of more elementary ice crystals, there may be 
many different Higgs bosons, and they may be composite objects made out of
constituents that are more elementary.

\subsection{How Heavy is the Higgs Snowflake?}

The direct search for the Standard Model Higgs boson at LEP established the
lower limit~\cite{LEPH}:
\begin{equation}
m_H \; > \; 114.4~{\rm GeV} .
\label{mHLEP}
\end{equation}
Moreover, as we saw in Fig.~\ref{fig:SMtests}, the precision electroweak data
are sensitive to both $m_t$ and $m_H$. Incorporating the current experimental
value $m_t = 173.1 \pm 1.3$~GeV, the best-fit value and 68\%
confidence-level range for the Higgs mass are~\cite{LEPEWWG}:
\begin{equation}
m_H \; = \; 89^{+35}_{-26}~{\rm GeV} .
\label{mHfit}
\end{equation}
The corresponding 95\% confidence-level upper limit is $m_H < 158$~GeV, or 185~GeV
if the direct limit (\ref{mHLEP}) is included.

\begin{figure}
\label{fig:TeVH}
  \includegraphics[height=.27\textheight]{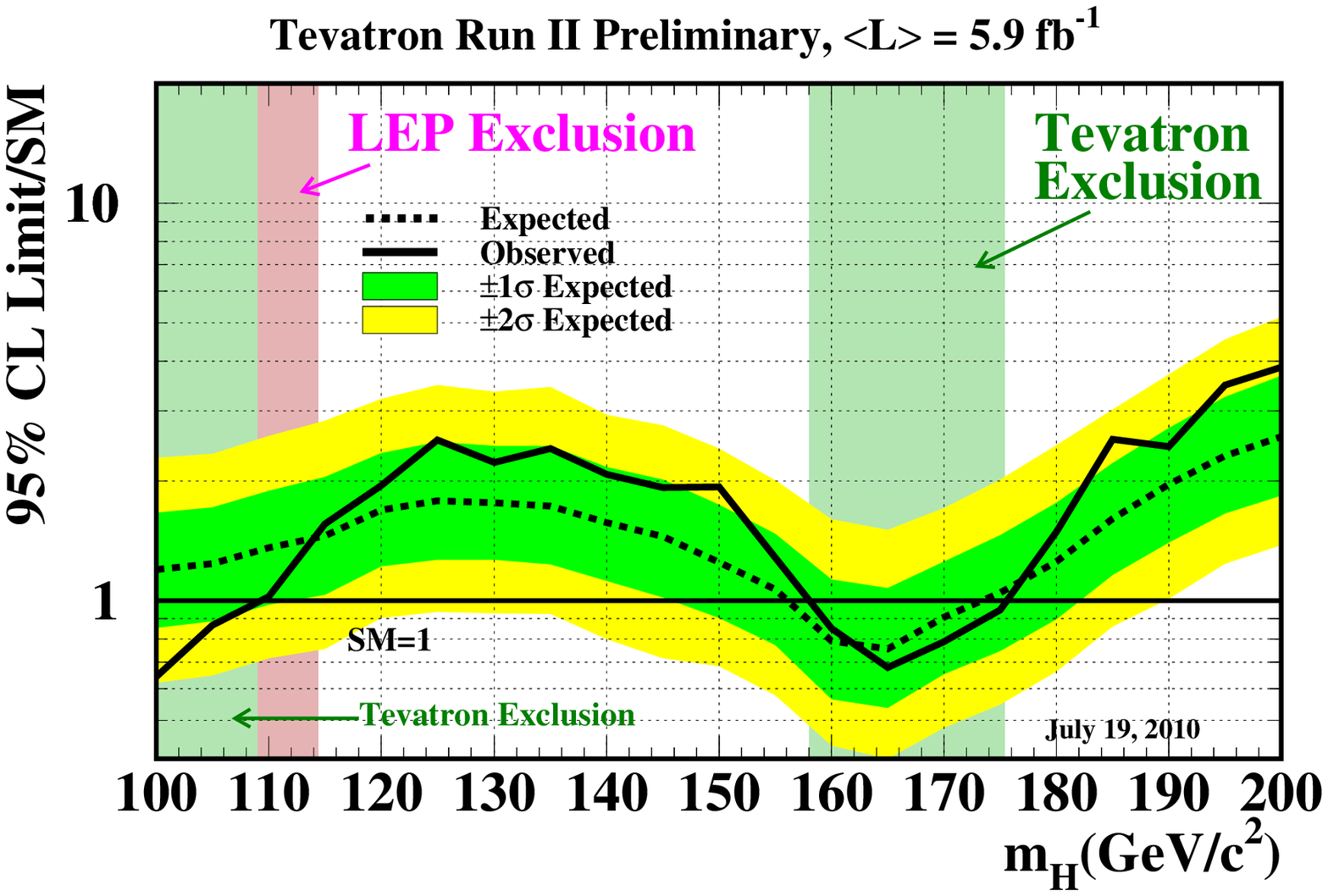}
  \includegraphics[height=.23\textheight]{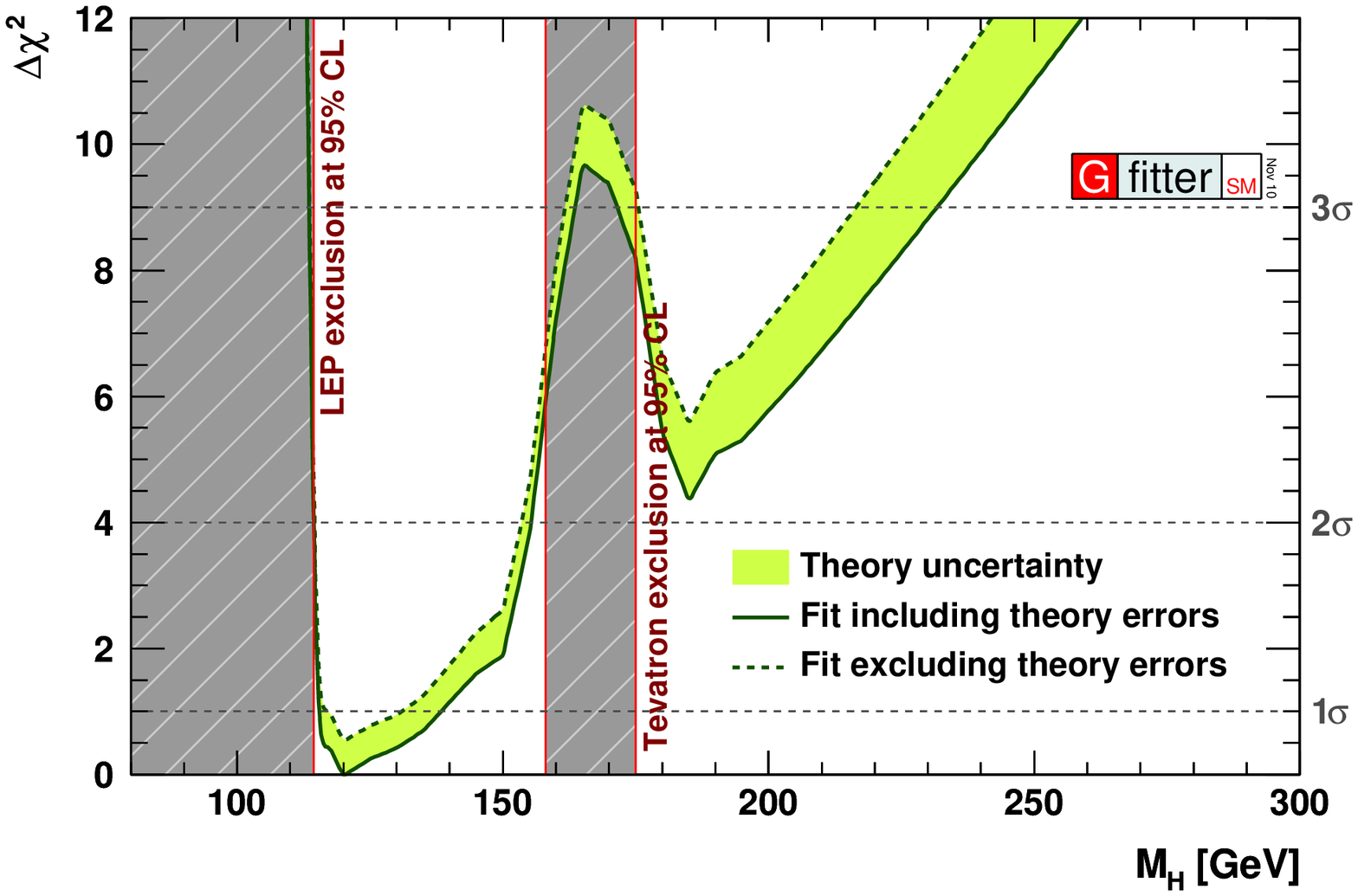}
  \caption{Searches for the Standard Model Higgs boson at the Tevatron (left) exclude the
  range $158~{\rm GeV} < m_H < 175$~GeV~\protect\cite{TeVH}. Combining the Tevatron search with the LEP
  search~\protect\cite{LEPH} and the precision electroweak data\protect\cite{LEPEWWG}, 
  one obtains (right) the global $\chi^2$ function that favours $m_H \sim 120$~GeV\protect\cite{GFitter}.}
\end{figure}

The direct experimental search for the Higgs boson is currently being led
by the Fermilab Tevatron, which has recently excluded the range~\cite{TeVH}:
\begin{equation}
\label{Tevatron}
158~{\rm GeV} \; < \; m_H \; < \; 175~{\rm GeV} ,
\end{equation}
as seen in the left panel of Fig.~\ref{fig:TeVH}. The right panel displays the result
of a combined $\chi^2$ analysis of the precision data with the direct searches at LEP
and the Tevatron. We see that the most likely value of the Higgs mass is $m_H \sim 120$~GeV~\cite{GFitter}.

\section{Dark Matter}

Astrophysicists and cosmologists tells us that there is five to ten times as much
invisible dark matter as the visible stuff out of which galaxies, stars, planets and people
are made~\cite{Dark}. The presence of this dark matter is felt gravitationally by visible matter,
whose velocities inside galaxies and clusters are much larger on average
than would be expected on the basis of the virial theorem and the density of
the visible matter itself. The galaxies and clusters need additional dark matter to keep 
them together, which might well be made out of massive neutral particles.
If these were once in thermal equilibrium with the visible matter in the early
Universe, one expects them to weigh less than about a TeV each, putting them
within reach of the LHC. There are many candidates in composite models, theories
with extra dimensions, etc., but here we concentrate on the lightest supersymmetric
particle (LSP)~\cite{EHNOS} as a prototype benchmark scenario, mentioning some others.

\section{Why I love SUSY}

Supersymmetry (SUSY) is the only symmetry that could unify matter particles and 
force particles. This is because it is unique in being able to relate particles spinning 
at different rates, such as the spin-0 Higgs boson, spin-$\frac{1}{2}$ matter particles
such as the electron and quarks, spin-1 intermediate bosons such as the photon, the
spin-$\frac{3}{2}$ supersymmetric partner of the graviton, called the gravitino, and the
spin-2 graviton itself~\cite{SUSY}. In addition, it would help fix particle masses~\cite{hierarchy} and unify the
fundamental forces~\cite{GUTs} and it predicts that the Higgs boson should be relatively light~\cite{SUSYH}, as
indicated by the precision electroweak data, as well as potentially providing the dark 
matter~\cite{EHNOS} postulated by the astrophysicists and cosmologists.

To see how SUSY could help the Higgs boson fix particle masses~\cite{hierarchy}, consider
loop corrections to the squared mass of the Higgs boson. Generic one-loop fermion 
and boson loops in the Standard Model
are each quadratically divergent, being $\propto \int^\Lambda d^4k/k^2$
where $\Lambda$ is a cut-off in momentum space, representing the maximum energy
scale up to which the Standard Model remains valid:
\begin{eqnarray}
\Delta m_H^2 & = & - \frac{y^2_f}{16 \pi^2} \left[ 2 \Lambda^2 + 
6 m_f^2 ln \left(\frac{\Lambda}{m_f} \right) \right] , \nonumber \\
\Delta m_H^2 & = & \frac{\lambda_s}{16 \pi^2} \left[ \Lambda^2 -
2 m_s^2 ln \left(\frac{\Lambda}{m_s} \right) \right] .
\label{quadratic}
\end{eqnarray}
Here $y_f$ denotes a Higgs-fermion-antifermion Yukawa coupling, and $\lambda_s$
is a quartic scalar coupling. If $\Lambda$ is of the same order as the grand unification
or Planck scale, these loop corrections are individually much greater than the possible
physical value of the Higgs mass. The presence of such quadratic divergences is {\it not}
incompatible with a light elementary Higgs boson, but it would seem quite
unnatural to obtain a light Higgs mass as the result of a cancellation between
the very cut-off-sensitive loop diagrams and a tree-level input contribution of the opposite sign. 
However, it is apparent that, since the fermion and scalar diagrams
have opposite signs, their quadratic divergences cancel if
\begin{equation}
\lambda_s \; = \; 2 y_f^2 .
\label{cancel}
\end{equation}
Remarkably this is exactly the relation between fermion and scalar couplings that
occurs in a supersymmetric theory, and the same relation cancels all quadratic
(and some logarithmic) divergences in all orders of perturbation theory~\cite{noren}. The residual
logarithms are not too large numerically if $\Lambda$ is of the same order as the 
grand unification or Planck scale, so SUSY restores the
naturalness of a light Higgs boson in a theory with light supersymmetric partners
of all the Standard Model particles. The supersymmetric particle mass scale
effectively replaces the upper cut-off $\Lambda$ on the validity of the Standard Model. 

Indeed, SUSY actually {\it predicts} a light Higgs boson, typically $m_H < 130$~GeV
in the minimal supersymmetric extension of the Standard Model~\cite{SUSYH}.

The appearance of supersymmetric particles would change the evolution of the
gauge couplings at larger energy-momentum scales. This is welcome, because
extrapolation of the measured gauge coupling strengths to high energies using
just the renormalization-group equations of the Standard Model reveals no energy
at which they would all be equal, making conventional unification impossible.
On the other hand, incorporating supersymmetric particles with masses $\sim 1$~TeV,
as suggested by the above naturalness argument, could bring the gauge couplings
together at some energy scale $\sim 10^{16}$~GeV, making possible unification of
the fundamental interactions~\cite{GUTs}. Further tests of unification would be made possible by
measuring the masses of the different supersymmetric particles~\cite{Zerwas}.

\section{Particle Cosmology}

The fact that the sky is dark at night tells us that the Universe cannot be in a strictly
steady state, and its current expansion was discovered by Hubble, who first observed the
redshifts in the light from other galaxies. The cosmic microwave background radiation,
emitted when atoms were first formed,
is evidence that the Universe was once about 1000 times hotter than it is today.
The cosmological abundances of light elements agree reasonably well with calculations based on
Big Bang nucleosynthesis (though see~\cite{Li}), and take us back to when the Universe was about $10^9$
times hotter than today. We believe that protons and neutrons were formed when the
Universe was about 100 times hotter still, and the LHC has recently been colliding
lead ions with energies of 2.76~GeV/nucleon in order to understand better the 
quark-gluon matter that filled the Universe before this epoch. Proton-proton collisions at the
LHC recreate quark and gluon collisions at energies similar to those typical of the
very early Universe when it was about 1000 times hotter still.
We believe that this is the epoch when particle masses appeared through the
Englert-Brout-Higgs mechanism, as described above. 

At least two major cosmological
mysteries may be resolved by the ability of LHC collisions to reach back to the
very early Universe. In typical models, the dark matter 
particles decouple from visible particles some time between the epochs of mass
generation and the transition from quark-gluon to hadronic matter. Additionally, it
is possible, e.g., in supersymmetric models, that the cosmological baryon asymmetry
was generated around the epoch of mass generation, as discussed next.

\section{The creation of matter}

Following the postulation of antimatter by Dirac and its discovery in the cosmic rays,
for over 30 years particle physicists thought that matter and antimatter particles
were exactly equal and opposite, having identical masses and opposite electric
charges. However, in 1964 an experiment revealed unexpectedly that some matter
and antimatter particles actually decay slightly differently, violating the combination of
charge conjugation and parity symmetries (CP), and also time-reversal symmetry (T). In 1967,
Sakharov~\cite{Sakharov} pointed out that such a matter-antimatter asymmetry combined with a departure
from thermal equilibrium during the expansion of the Universe could enable a difference
between the cosmological abundances of matter and antimatter to be created.
If such an excess of matter particles was created, around the epoch of the transition from 
quark-gluon matter to hadronic matter, all the particles of antimatter would have
annihilated with matter particles, leaving a surplus of the latter to survive into the Universe today.

Then, in 1973 Kobayashi and Maskawa showed that CP and T violation could be
accommodated in the Standard Model with six quarks, and this paradigm has
been established by many subsequent experiments~\cite{CPV}. Could this mechanism be responsible 
for the creation of the matter in the Universe? Apparently not, because no strong breakdown
of thermal equilibrium is expected to have occurred in the Standard Model, and the amount of
Kobayashi-Maskawa CP violation seems inadequate.

However, many theories beyond the Standard Model, including SUSY, contain extra
sources of CP violation and mechanisms for matter creation, and some of these could
have created a matter-antimatter asymmetry at the epoch of the transition that generated
particle masses~\cite{SUSYBG}. Such theories are susceptible to experimental tests at the LHC, and one of
its experiments, LHCb, is dedicated to the study of CP violation and rare $B$ decays that
might cast light on the creation of matter - though other realizations of Sakharov's idea
would involve physics at earlier epochs beyond the direct reach of the LHC.

\section{Towards a Theory of Everything?}

Unifying the fundamental interactions was Einstein's dream in his latter decades,
and extra dimensions were among the ideas he explored. They also play essential
roles in many contemporary scenarios for unification and quantum gravity, e.g., in
the context of string theory. In fact, string theory seems to require both extra dimensions
and SUSY, though our present understanding is insufficiently advanced to calculate
the energy scales at which they might appear. In some scenarios with extra dimensions,
gravity becomes strong at the TeV scale, and microscopic black holes might be
fabricated in quark and gluon collisions at the LHC~\cite{LHCBH}. If so, their decays would provide
wonderful laboratories for probing theories of quantum gravity, e.g., by measuring
the grey-body factors of Hawking radiation into different particle species~\cite{Webber}.

\section{The LHC Physics Haystack(s)}

Why has the LHC not discovered anything yet?
Cross sections for heavy particles typically scale as $1/M^2$, and many, e.g., the Higgs
boson, have cross sections suppressed by powers of small couplings. For these reasons,
their cross sections are much smaller than the total cross section, which is ${\cal O}(1/m_\pi^2)
\sim 1/(100~{\rm MeV})^2$. Therefore, cross sections for new physics are typically a trillionth of the
total cross section. Since many new particle signatures, e.g., for the Higgs boson,
are accompanied by large backgrounds, many events may be needed to establish a signal.
Looking for new physics at the LHC is like looking for a needle in $\sim 100,000$ haystacks!
At the time of writing, the LHC experiments have each accumulated just a few trillion events,
so it should not be surprising that they have not yet discovered new physics. Nevertheless,
already the LHC has established some of the strongest limits on new physics, as discussed below,
and the number of LHC collisions may increase by a factor $\sim 100$ in the coming year,
putting it firmly in the discovery business.

\begin{figure}[h!]
  \includegraphics[height=.25\textheight]{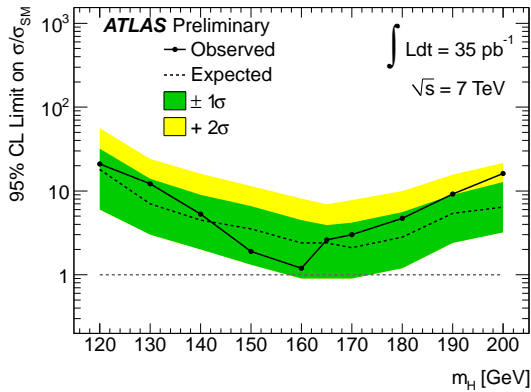}
   \caption{The result of an initial ATLAS search for the Higgs boson in the $H \to WW$ channel, showing
  the expected signal rate, relative to the SM rate, that is excluded at the 95\% CL~\protect\cite{Ketevi}.}
  \label{fig:Ketevi}
\end{figure}

\section{The Search for the Higgs Boson}

\begin{figure}
\hspace{-5.2cm}
  \includegraphics[height=.2\textheight, angle=270]{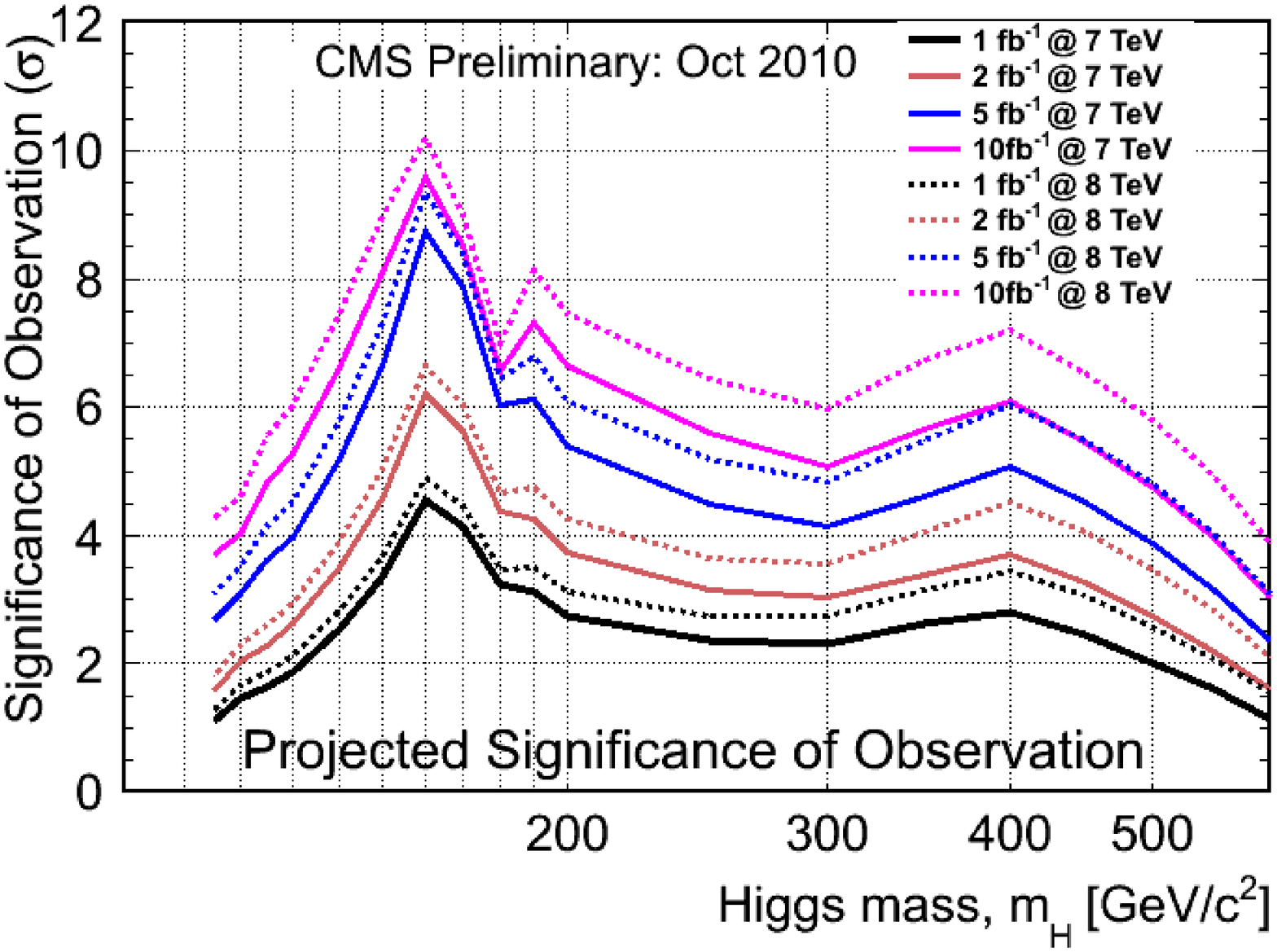}
  \caption{The expected sensitivities of the ATLAS (upper) and CMS (lower) experiments
  at the LHC for observing the Standard Model Higgs, as a function of the integrated luminosity
  and centre-of-mass energy~\protect\cite{ATLASH,CMSH}.}
  \label{fig:LHCH}
\end{figure}

In the range of Higgs masses below 150~GeV, which currently seems the most plausible,
several different Higgs production and decay modes may contribute to the search for the Higgs boson at 
the LHC, including $gg, W^+ W^- \to H \to \gamma \gamma, \tau \tau, W^+ W^-$ and $Z^0 Z^{0*} \to 4$ leptons.
The result of an initial ATLAS search for the Higgs boson are shown in Fig.~\ref{fig:Ketevi}: already with only 35/pb
of data analyzed, the LHC upper limit on Higgs production approaches the Standard Model expectation and the
Tevatron limit. The latest estimates by ATLAS~\cite{ATLASH} and CMS~\cite{CMSH} 
of their likely future sensitivities to a Standard Model Higgs
boson are shown in Fig.~\ref{fig:LHCH}, for various assumptions about the available LHC
integrated luminosity and centre-of-mass energy. It is now planned to extending the present run into 2011,
operating at 7~TeV this year but maybe increasing the LHC energy to 8~TeV in 2012,
which should provide good prospects of discovering (or excluding) a Standard Model
Higgs boson at any mass up to $\sim 600$~GeV. In parallel, it had been proposed to
extend the Tevatron run for three years, offering the prospect of discovering a light
Higgs boson {\it via} complementary production and decay channels, such as $W + H, H \to
{\bar b}b$, providing valuable additional science. Unfortunately, this proposal has not been accepted~\cite{B2S}. Nevertheless, the question of the origin of particle masses may soon be answered.

\begin{figure}[t!]
  \includegraphics[height=.25\textheight]{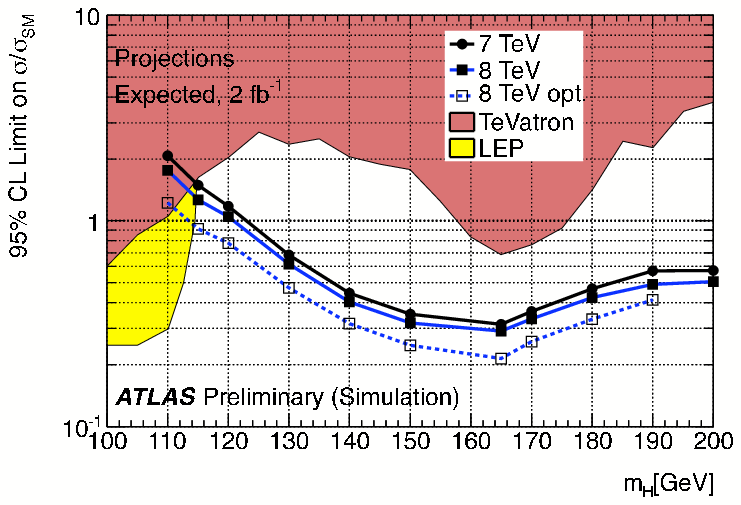}
  \end{figure}

The stakes in the Higgs search are high. The answer to the mass question will tell us
how the symmetry between different particles is broken, and whether there is an 
elementary scalar field - something which has never been seen and would surprise
many theorists. The existence and mass of the Higgs boson will also foretell the fate 
of the Standard Model at high energies~\cite{EEGHR}, thereby establishing the framework for possible
unified theories. It will also tell us whether and how mass appeared when the Universe 
was a picosecond old, and may indicate whether the Higgs could have played a role in
creating the matter in the Universe. The existence of a Higgs boson could have other
cosmological implications. For example, many models of inflation postulate a similar
elementary scalar field (or even the Higgs field itself~\cite{Hinflation}) to explain the size and age of
the Universe. Moreover, the Higgs has the potential to contribute
$\sim 10^{60}$ times more dark energy than what is observed, and measurements of the Higgs boson
may cast light on the problem of dark energy. 

\section{The Search for SUSY}

\subsection{Supersymmetric models}

In the following we discuss the prospects for SUSY searches in the context
of the minimal supersymmetric extension of the Standard Model (the MSSM), in which known
particles are accompanied by spartners with spins differing by $\frac{1}{2}$, and there are
two Higgs doublets, with a coupling $\mu$, and v.e.v.'s in the ratio of $\tan \beta$.
In addition, there are unknown parameters that characterize supersymmetry breaking,
namely soft scalar masses $m_0$, spin-$\frac{1}{2}$ gaugino masses $m_{1/2}$, 	trilinear 
soft supersymmetry-breaking couplings $A_\lambda$, and a bilinear soft coupling $B_\mu$.
The MSSM has over 100 parameters, too many for practical phenomenology until many
more experimental constraints become available, e.g., from the LHC. In the mean time, it is
often assumed that the scalar and gaugino masses are universal, and likewise the
trilinear couplings. This is consistent with experimental data and measurements of
rare flavour-changing processes, which suggest a super-GIM mechanism~\cite{EN} as would be provided
by universal $m_0$ parameters for the squarks and sleptons in different generations
but with the same quantum numbers~\cite{BG},
and GUTS, which would link the $m_0$ parameters of squarks and sleptons in the same GUT multiplet
and possibly also the $m_{1/2}$ parameters for the SU(3), SU(2) and U(1) gauginos. This is
the simplified phenomenological framework known as the constrained MSSM (the CMSSM),
which has just 4 variables and the sign of $\mu$ as parameters.
Unfortunately, there is no strong motivation for it from fundamental theory such as strings,
and one may consider alternatives.

Generalizing the CMSSM, one may note that none of the arguments in the previous
paragraph give any reason why the soft supersymmetry-breaking contributions to the masses
of the two Higgs doublets should be universal, and one may consider non-universal Higgs
mass models in which they are either equal (the NUHM1) or unequal (the NUHM2). Alternatively,
one may consider more constrained models, such as minimal supergravity (mSUGRA), which fixes the
gravitino: $m_{3/2} = m_0$ and imposes the relation $B_\mu = A_\lambda - m_0$. One may also
consider an intermediate, very constrained model (the VCMSSM) in which the relation
$B_\mu = A_\lambda - m_0$ is retained but the gravitino mass relation is dropped. In the following,
we will compare the prospects for SUSY searches in the CMSSM, NUHM1, VCMSSM and mSUGRA.

\subsection{Candidates for dark matter}

Many supersymmetric models have a multiplicatively-conserved $R$-parity: $R = (-1)^{2S - L + 3B}$,
where $S, L$ and $B$ denote the spin, lepton and baryon numbers, respectively~\cite{Fayet}. In such models,
heavier sparticles are condemned to be produced in pairs and to decay into lighter ones, so as to 
conserve $R$-parity, and the lightest sparticle (LSP) is stable, as it has no allowed decay mode. Hence, 
it could lurk around today as a relic from the Big Bang, and constitute the dark matter. Other models
such as scenarios with universal extra dimensions and composite models may have analogous dark matter
candidates, the LKP~\cite{LKP} and LTP~\cite{LTP}, respectively, that are benchmarked by discussing the LSP.

The LSP (LKP, LTP) cannot be charged or have strong interactions, as otherwise it would bind to
conventional particles forming anomalous heavy `nuclei' that have not been seen. {\it A priori}, 
weakly-interacting LSP (LKP) candidates in the MSSM (universal extra dimension scenario)
include the supersymmetric partner (Kaluza-Klein excitation) of either (i) some neutrino $\tilde \nu$ ($\nu_{KK}$), 
or (ii) a mixture of the neutral SU(2) and U(1) gauginos and Higgs bosons, namely the lightest neutralino $\chi$ 
($V_{KK}$), or (iii) the gravitino. In the supersymmetric framework, the $\tilde \nu$ is apparently excluded by 
a combination of LEP data and direct searches for astrophysical dark matter, and in the following we focus 
on the lightest neutralino $\chi$~\cite{EHNOS}, 
whilst recognizing that the gravitino is also a valid possibility that would
have distinctive signatures at the LHC but be very difficult to detect in any astrophysical context. 

\subsection{Constraints on SUSY}

\begin{figure}
  \includegraphics[height=.3\textheight]{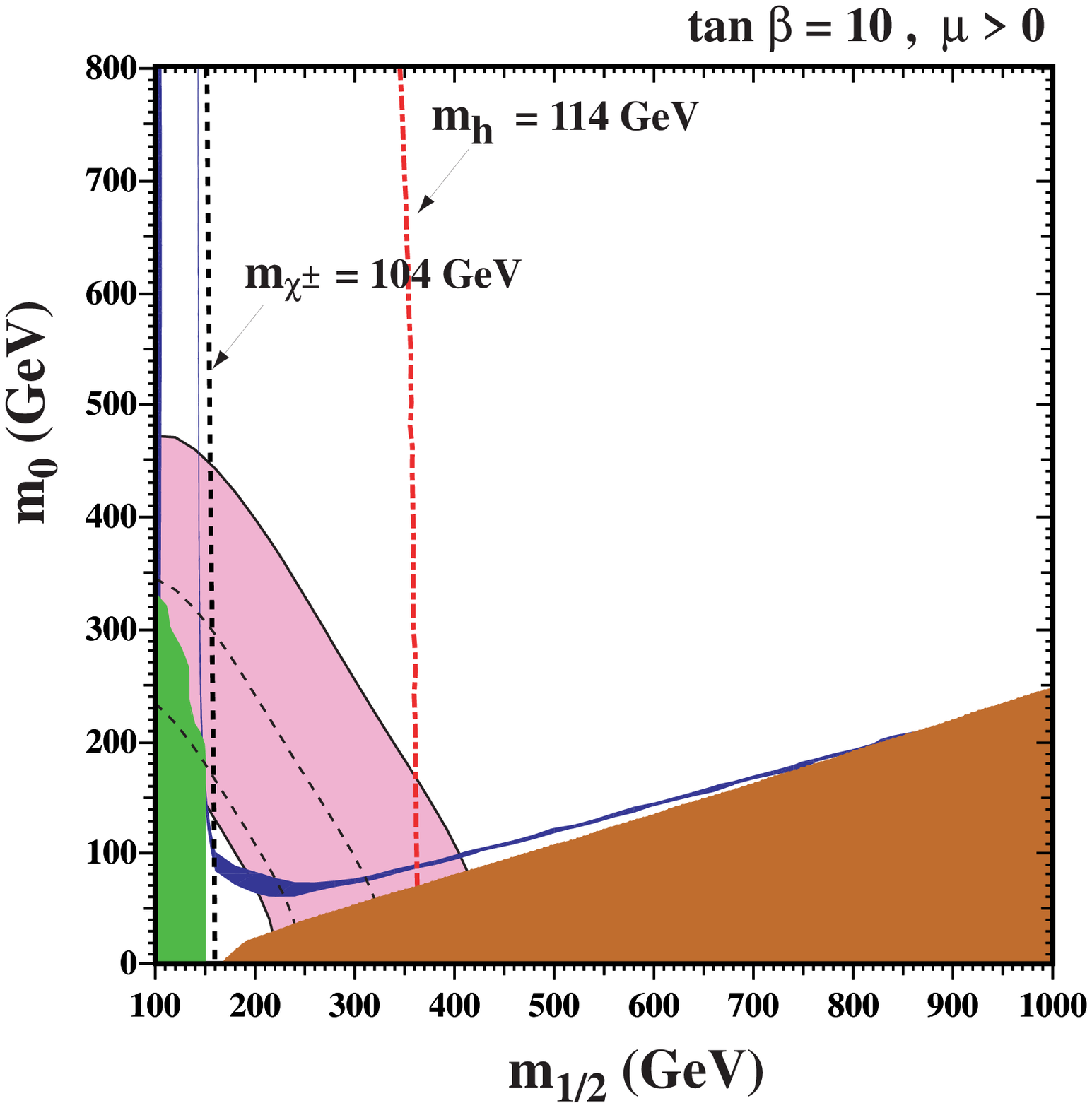}
  \includegraphics[height=.3\textheight]{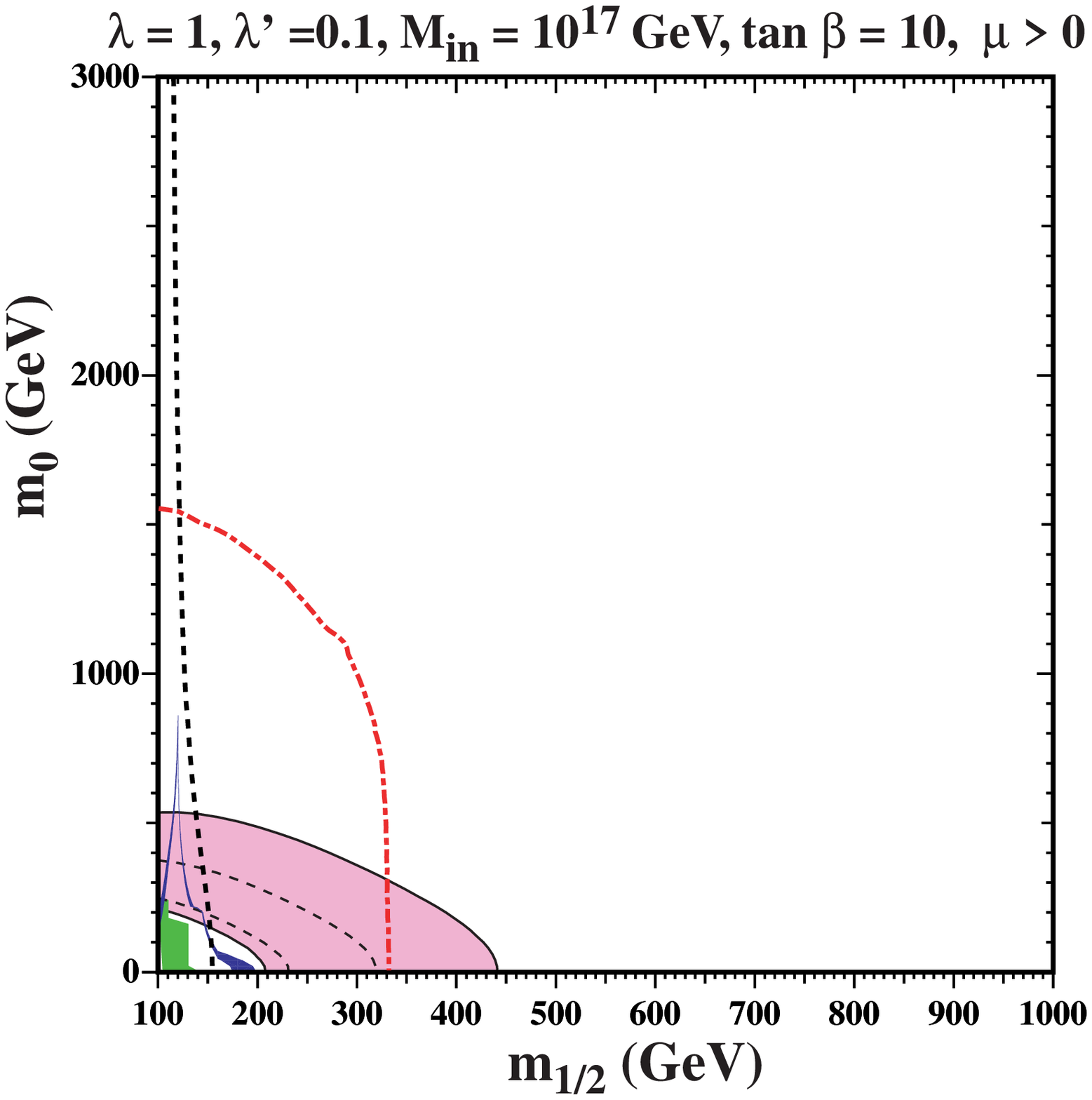}
  \caption{The $(m_{1/2}, m_0)$ planes for the CMSSM with $\tan \beta = 10$ and
  $\mu > 0$ assuming (left) universality at the grand unification 
  scale $\sim 10^{16}$~GeV~\protect\cite{EOSS} and (right)
 assuming universality at $10^{17}$~GeV~\protect\cite{EMO}. The near-vertical (red)
dot-dashed lines are the contours $m_h = 114$~GeV~\protect\cite{LEPH}, 
and the near-vertical (black) dashed
line is the contour $m_{\chi^\pm} = 104$~GeV~\protect\cite{LEPsusy}.  The medium (dark
green) shaded region is excluded by $b \to s
\gamma$, and the dark (blue) shaded areas make up the region favoured by determinations of the cosmological
dark matter density~\protect\cite{WMAP}. The dark
(brick red) shaded region is excluded because there the LSP would be the charged lighter stau slepton. The
region favoured by the E821 measurement of $g_\mu -2$ at the 2-$\sigma$
level, is shaded (pink) and bounded by solid black lines, with dashed
lines indicating the 1-$\sigma$ ranges~\protect\cite{g-2}.}
  \label{fig:tb10}
\end{figure}

The classic collider signature for any dark matter candidate is missing transverse momentum, as inferred from an
imbalance in the transverse energy, and in the neutralino LSP scenario the absence of such a signature at LEP 
and the Tevatron collider implies that most sparticles must weigh $> 100$~GeV and squarks and gluinos must
weigh $> 400$~GeV. The absence of the Higgs boson at LEP and the consistency of $B$ decays such as $b 
\to s \gamma$ and $B_s \to \mu^+ \mu^-$ also impose important constraints on SUSY. The most tangible
positive indication for SUSY is the cosmological density of dark matter. Since the density is known with a
precision of a few percent~\cite{WMAP}, so also is some combination of SUSY model parameters in any given 
scenario. The left panel of Fig.~\ref{fig:tb10} shows a compilation of constraints in the $(m_{1/2}, m_0)$ plane for 
the CMSSM with $\mu > 0$ and $\tan \beta = 10$, assuming that the dark matter is composed 
of neutralinos $\chi$ and that the universality of the CMSSM applies at an input grand unification
scale $\sim 10^{16}$~GeV~\cite{EOSS}. In addition to the phenomenological constraints mentioned above, this 
figure also shows the region of parameter space excluded because the LSP is charged. 
Finally, also displayed is the region
that would be favoured if one interprets the apparent discrepancy between experiment~\cite{g-2}
and the Standard Model calculation of the anomalous magnetic moment of the muon, 
$g_\mu - 2$, as being due to SUSY. The validity of this interpretation is still contested~\cite{Davier}, 
so we also discuss below the implications of dropping it. The right panel of Fig.~\ref{fig:tb10} shows a 
similar compilation of constraints in the $(m_{1/2}, m_0)$ plane, this time assuming that the universality of
the CMSSM applies at an input scale of $10^{17}$~GeV~\cite{EMO}, revealing a rather different
picture! In the following, we assume CMSSM universality at the grand unification scale.

\begin{figure}[t!]
  \includegraphics[height=.21\textheight]{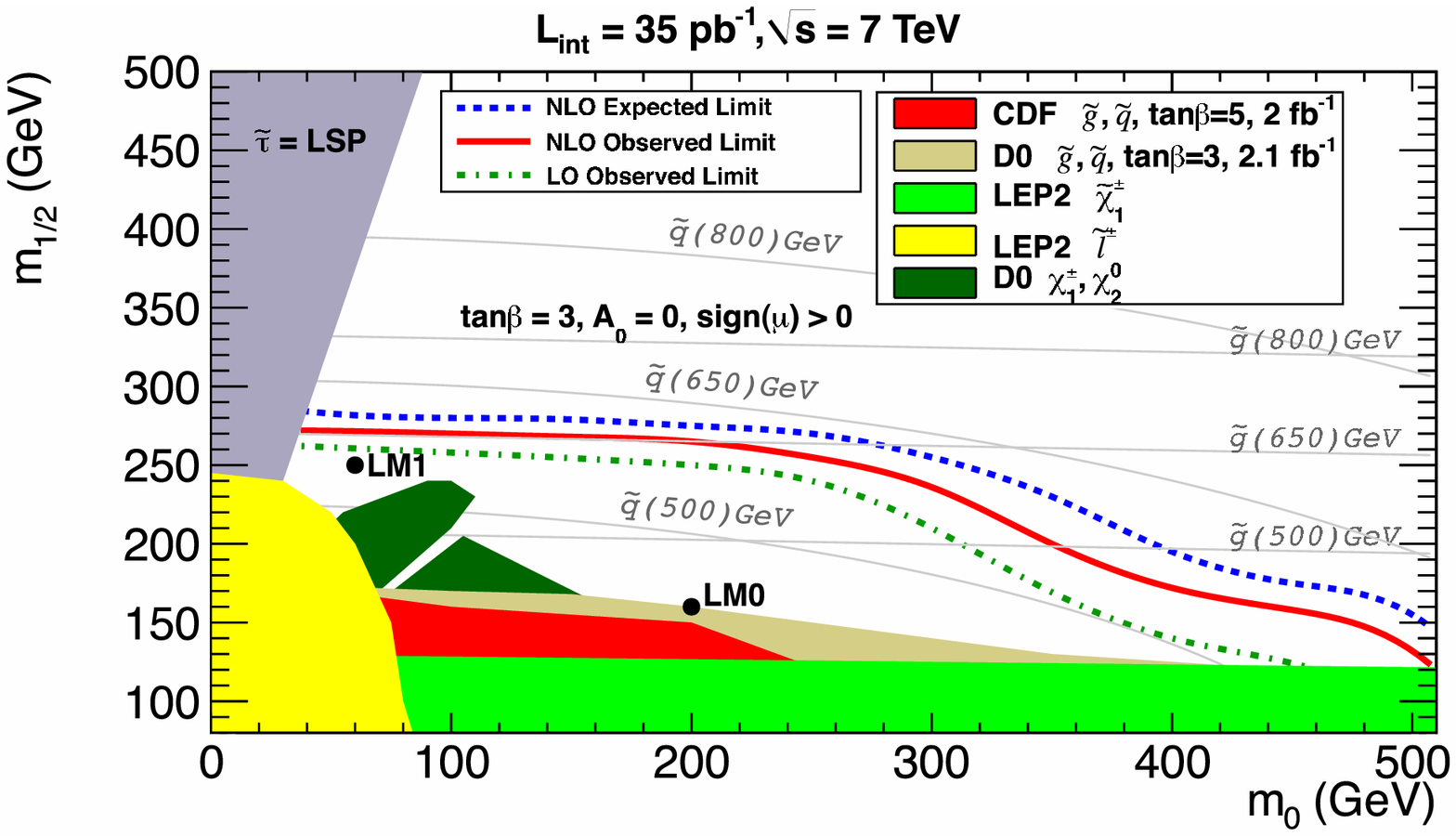}
\includegraphics[height=.20\textheight]{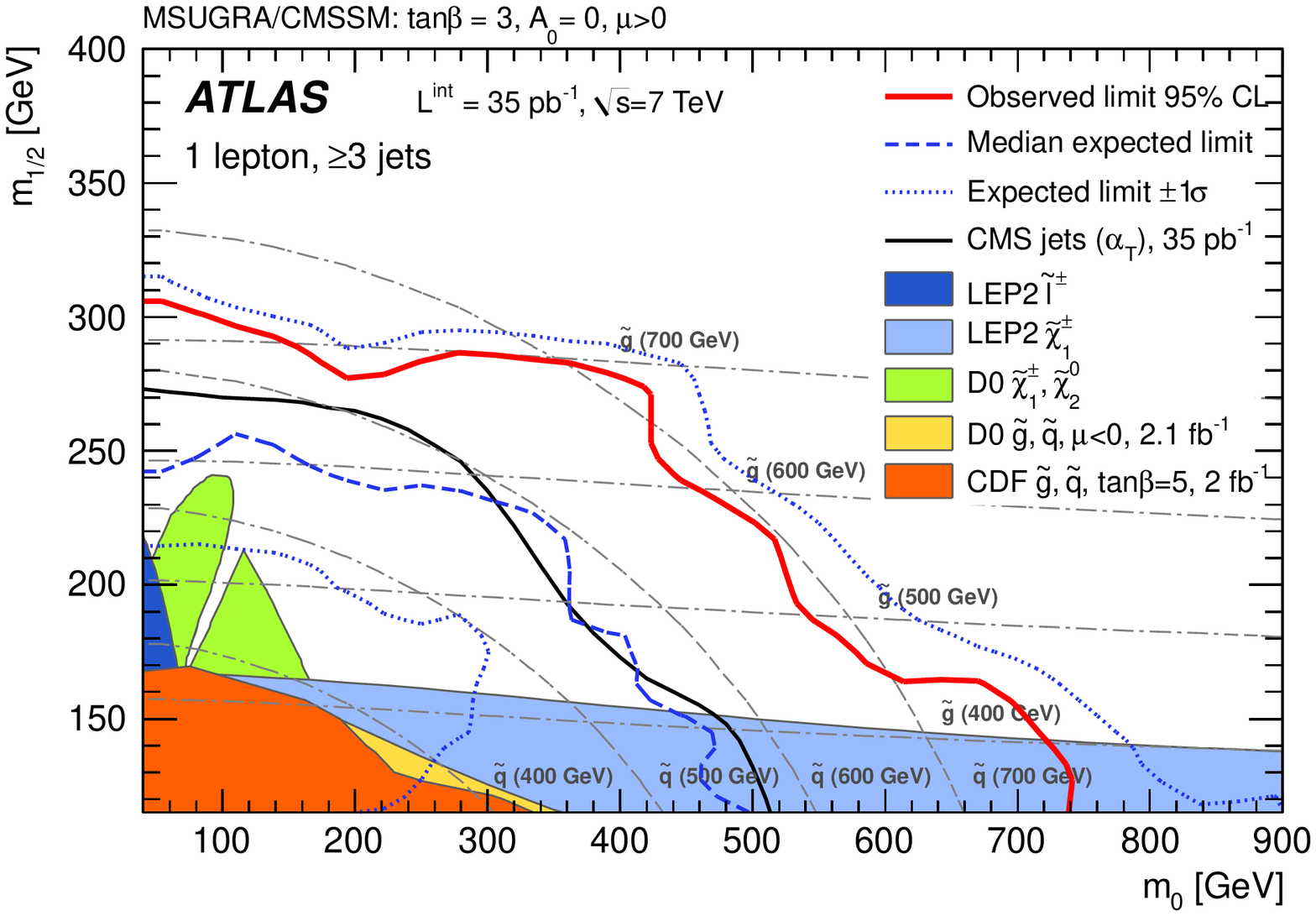}
  \caption{The exclusions in the $(m_0, m_{1/2})$ plane from (left) the initial CMS search
  for jets + missing transverse energy events with 35/pb of analyzed data at 7 TeV in the centre of 
  mass~\protect\cite{CMSlimit}, and (right) the initial
  ATLAS search for lepton + jets + missing transverse energy events~\protect\cite{ATLASlimit}.}
  \label{fig:observed}
\end{figure}

\subsection{Global supersymmetric fits}

We now present some results from frequentist supersymmetric 
fits to the parameters of the CMSSM, NUHM1, VCMSSM
and mSUGRA~\cite{MC4}, incorporating contributions from
all the above constraints to total likelihood function,
also including the constraints provided 
by the initial LHC searches for SUSY reported in~\cite{CMSlimit,ATLASlimit}
shown in Fig.~\ref{fig:observed}, as discussed in~\cite{MC5}.

\begin{figure}[t!]
  \includegraphics[height=.25\textheight]{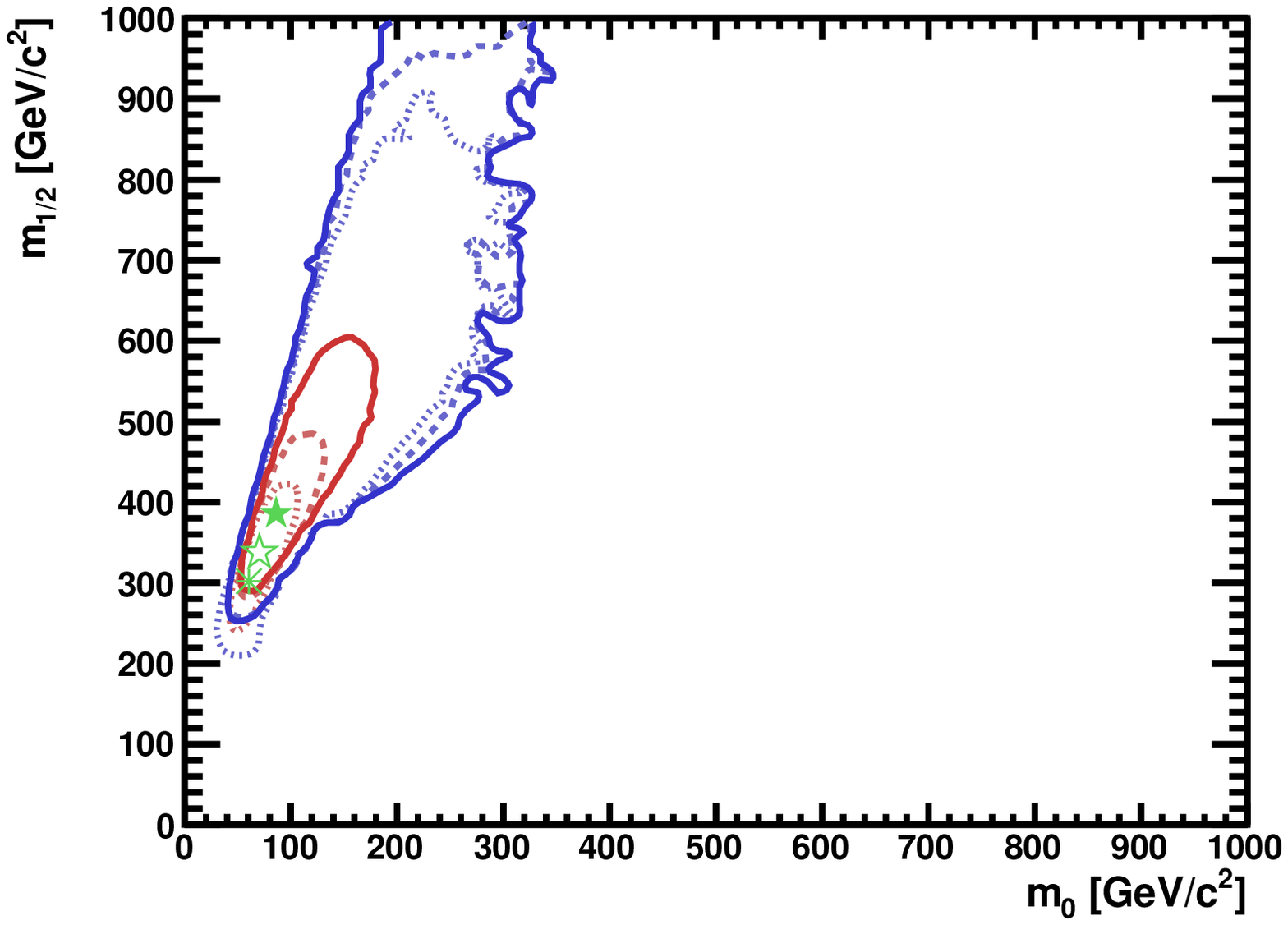}
  \includegraphics[height=.25\textheight]{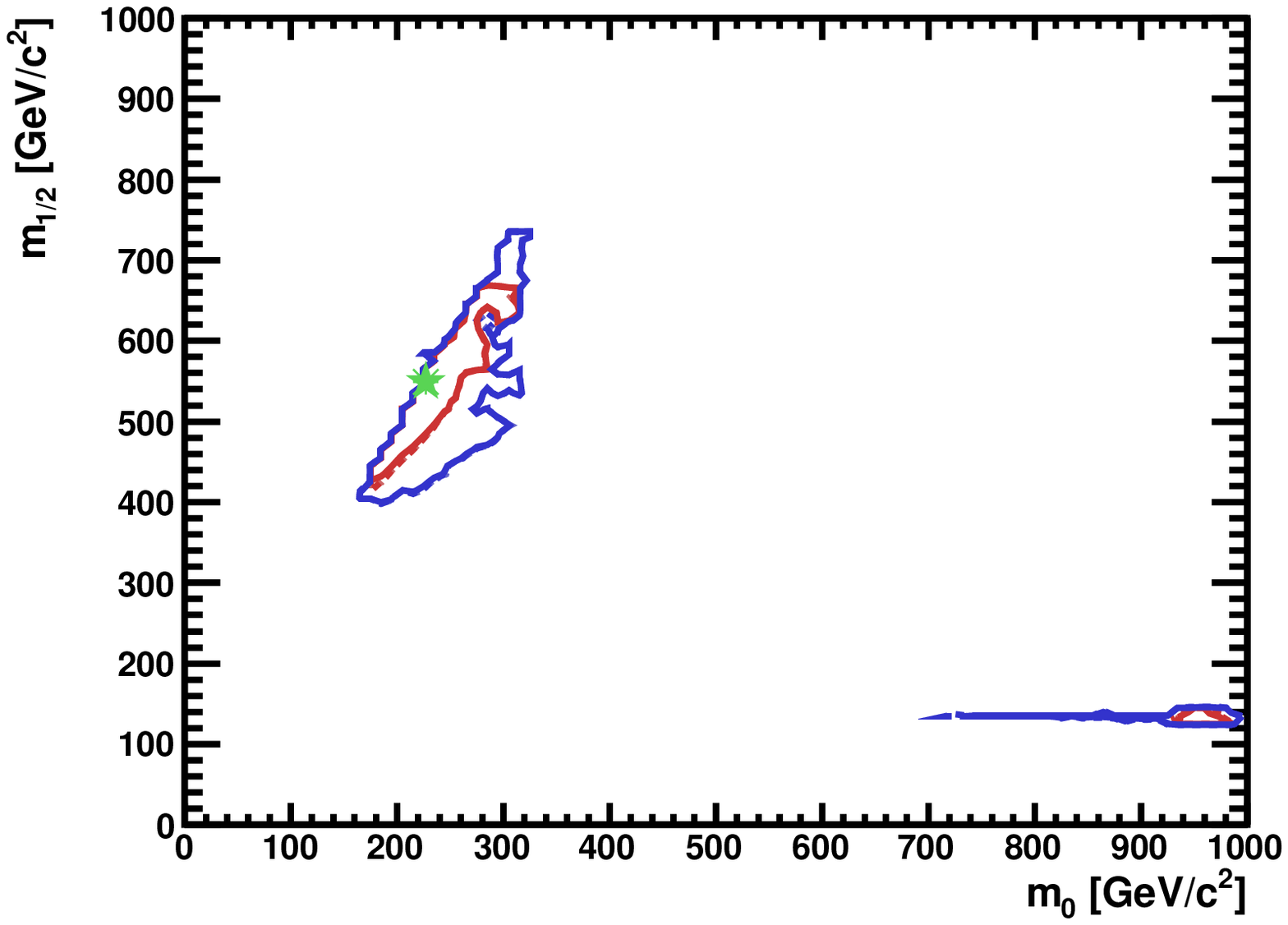}
  \caption{The $(m_0, m_{1/2})$ planes in the CMSSM (upper left),
NUHM1 (top right), VCMSSM (lower left) and mSUGRA (lower right).
In each panel, we show  the 68 and 95\%~CL contours (red and blue,
respectively) found in a frequentist analysis of the available 
constraints~\protect\cite{MC4,MC5}, both before applying the LHC constraints
(dotted lines) and after applying the 
CMS~\protect\cite{CMSlimit} constraint (dashed line) and the ATLAS constraint~\protect\cite{ATLASlimit}
(solid line). Also shown as (green) snowflakes, open and full stars are the
best-fit points in each model.}
  \label{fig:allm0m12}
\end{figure}

Fig.~\ref{fig:allm0m12} displays the $(m_0, m_{1/2})$ planes for these models, showing the
best-fit points as well as the regions favoured at the 68 and 95\% CL. The differences between the
dotted, dashed and solid lines illustrate the impact of the initial LHC constraints from the CMS 
and ATLAS Collaborations
shown in the left and right panels of Fig.~\ref{fig:observed}~\cite{CMSlimit,ATLASlimit}, 
respectively, in which no significant excess of
SUSY-like events were reported. 

\begin{figure}[t!]
  \includegraphics[height=.25\textheight]{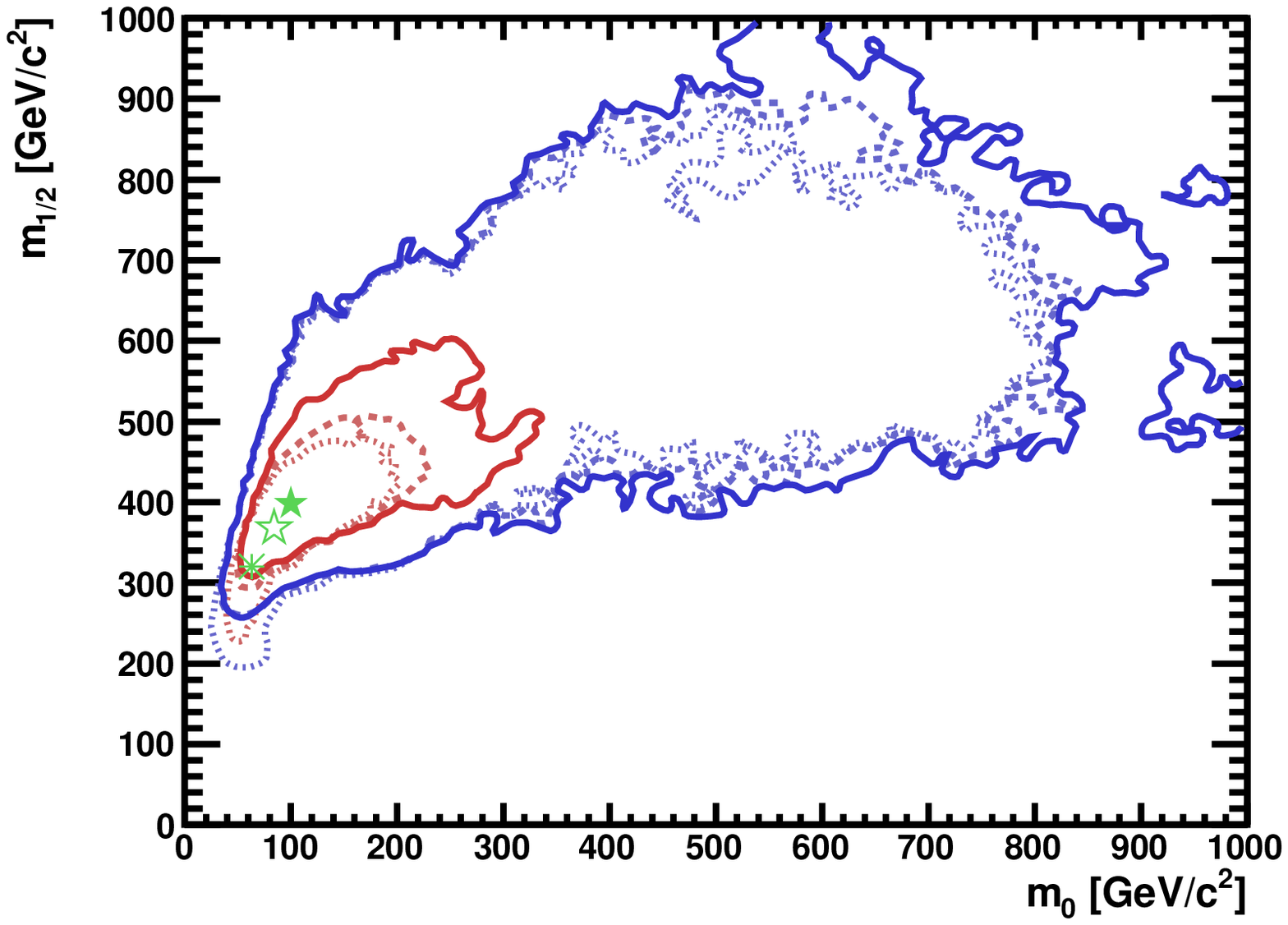}
  \includegraphics[height=.25\textheight]{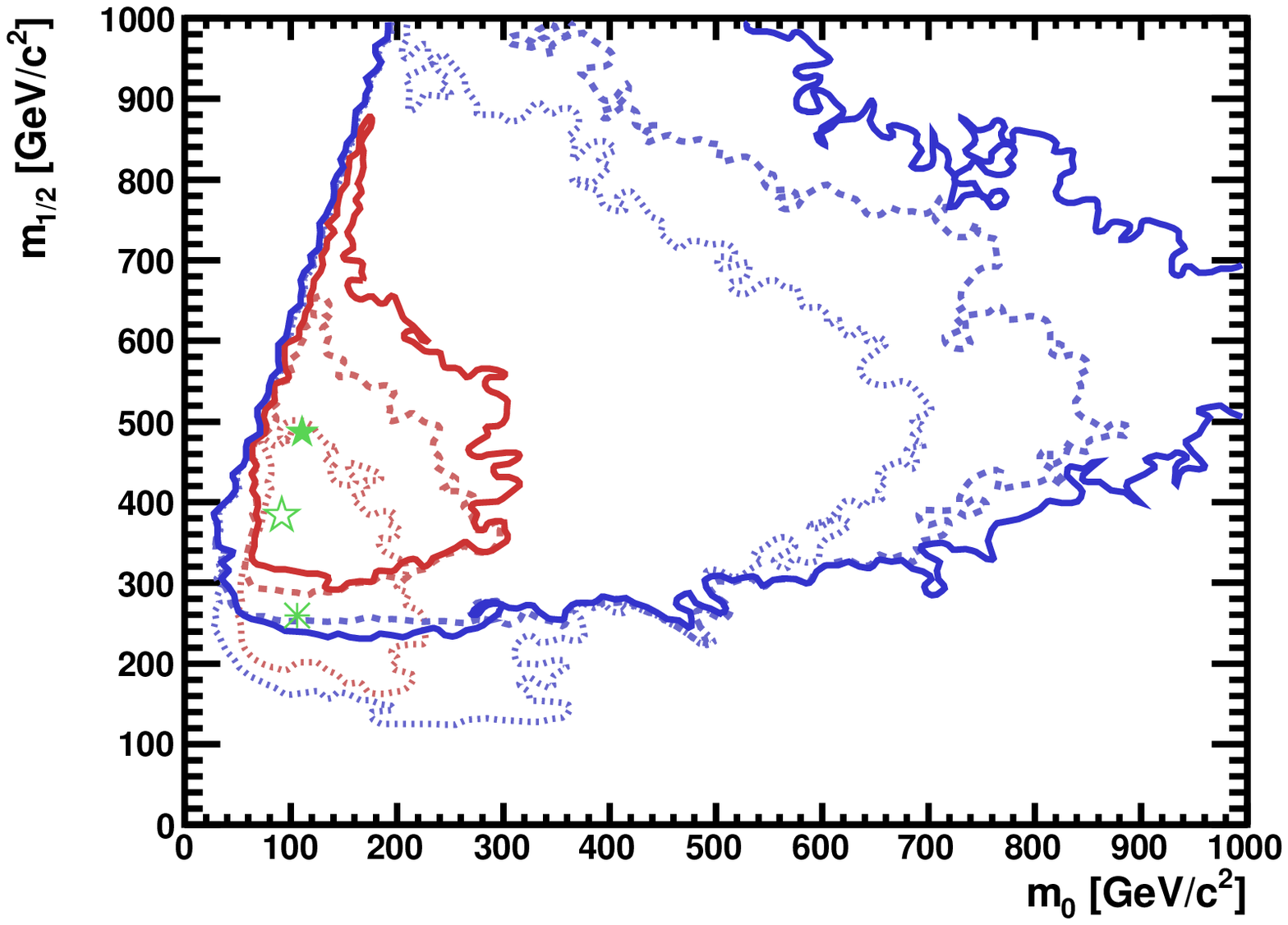}
  \end{figure}

\begin{figure}
  \includegraphics[height=.28\textheight]{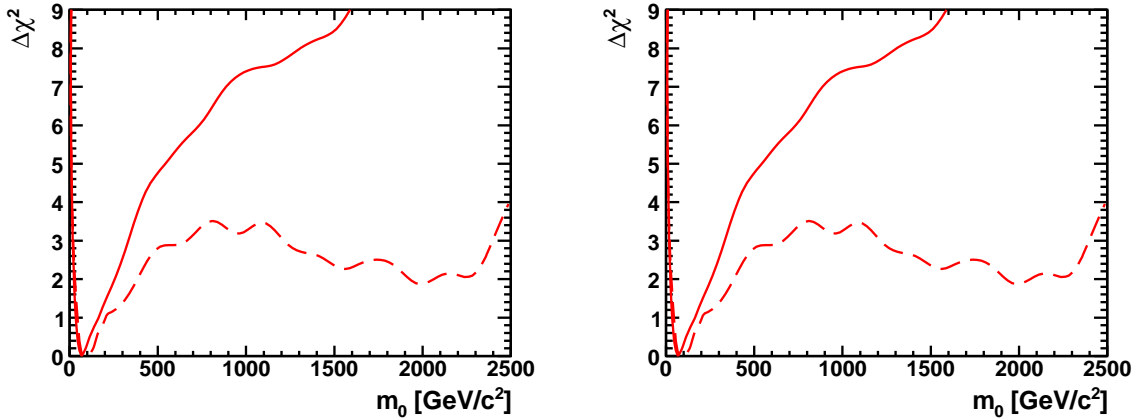}
   \includegraphics[height=.28\textheight]{m0CMSSMnog2}
  \caption{The likelihood functions for $m_0$ in (left) the CMSSM and (right) the NUHM1. 
  The $\chi^2$ values including (excluding) the $g_\mu - 2$ constraint are 
  shown as the solid (dashed) curves~\protect\cite{MC3}.}
  \label{fig:nog-2}
\end{figure}

Comparing with the
expected reaches for SUSY detection at the LHC~\cite{ATLASsusy,CMSsusy}, 
there should be good prospects for discovering SUSY in the near future. It
should be stressed, however, that these conclusions depend quite critically on
the $g_\mu - 2$ constraint: as seen in Fig.~\ref{fig:nog-2} for the CMSSM case
before applying the LHC constraints,
the other data show only a slight preference for light sparticles, e.g., {\it via} the
measurement of $m_W$ shown in the right panel of Fig.~\ref{fig:SMtests}. The
single-variable $\chi^2$ for some sparticle masses and other observables are
shown in Fig.~\ref{fig:predictions}. We note that the gluino is expected to weigh $<
1.5$~TeV in all these models, potentially within the reach of the LHC in 
2011/12, that the Higgs boson is predicted to weigh between 115 
and 120~GeV (the curves shown have a theoretical uncertainty estimated
at $\pm 3$~GeV), and that $B_s \to \mu^+ \mu^-$ decay may occur at a 
rate measurably different from the Standard Model prediction, particularly in the NUHM1.
LHCb may attain sensitivity close to this prediction also in 
2011/12.

\subsection{Searches for SUSY dark matter}

Several searches to search for supersymmetric dark matter have been proposed, principally with the lightest
neutralino $\chi$ in mind. These include searches for $\chi \chi$ annihilations in the galactic halo into
antiprotons, positrons, etc., that could be detected among the cosmic rays. Another possibility is to
look for annihilations into $\gamma$ rays in the galactic centre. A third possibility is to look for
annihilations into energetic neutrinos in the core of the Sun or Earth. Most promising may be
to search directly for $\chi$ scattering on nuclei in the laboratory.

\begin{figure}
  \includegraphics[width=.5\textwidth]{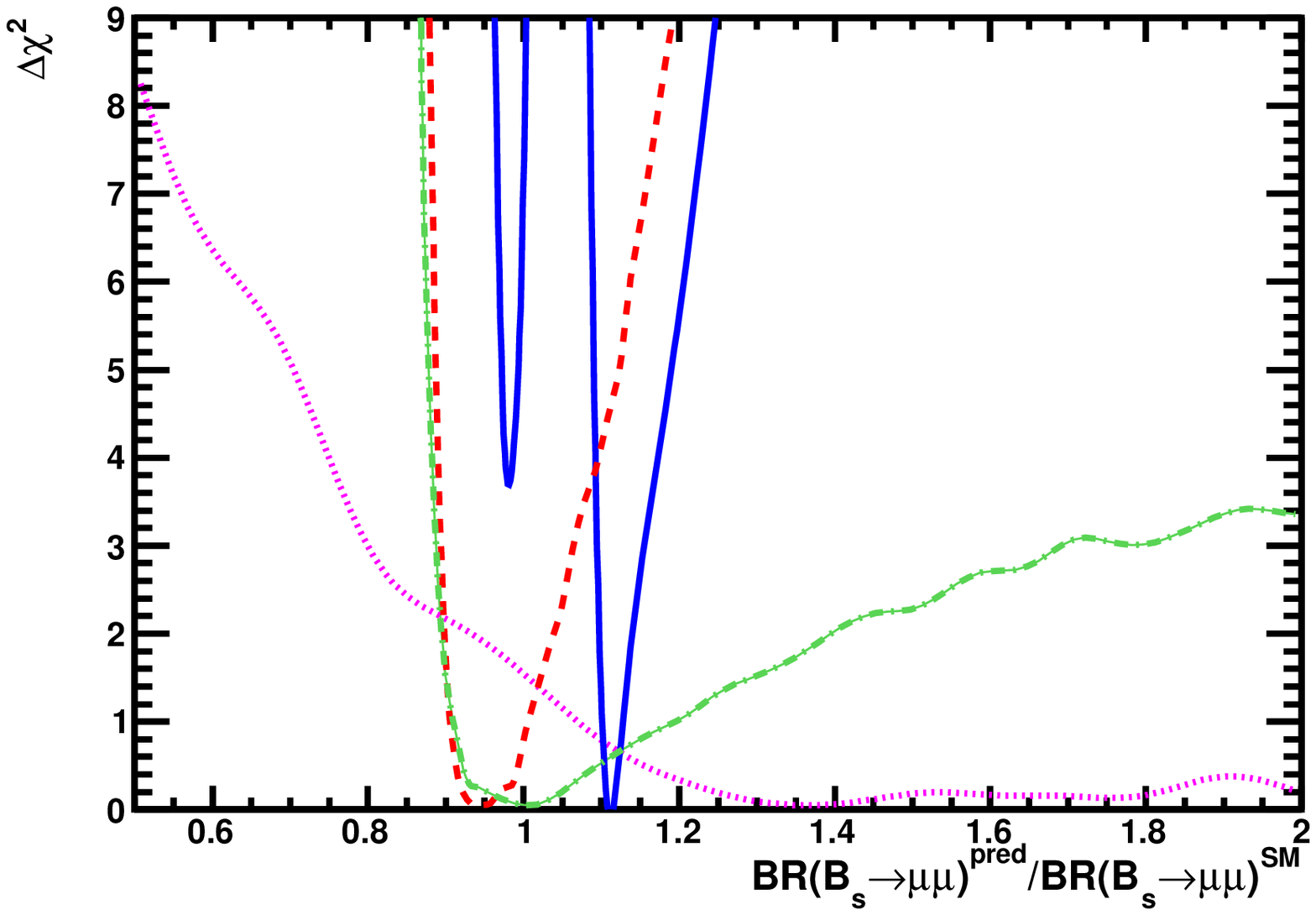}
  \includegraphics[width=.5\textwidth]{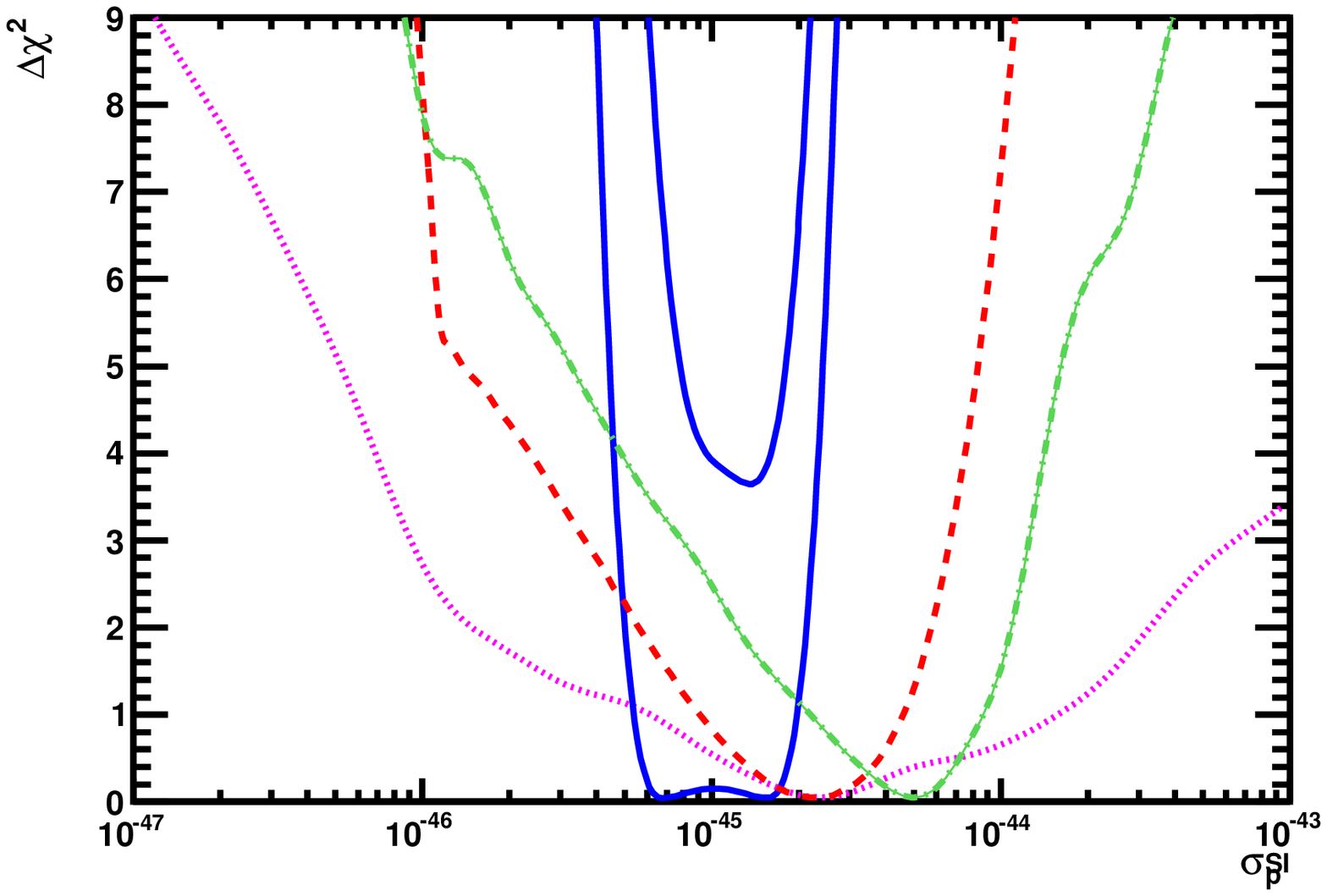}
  \caption{The likelihood functions for (upper left) $m_{\tilde g}$, (upper right) $m_H$
  (lower left) BR($B_s \to \mu^+ \mu^-$) and (lower right) the spin-independent $\chi - p$
  scattering cross section for the CMSSM (green dash-dotted lines), the NUHM1 (purple dotted lines), 
the VCMSSM (red dashed lines),  and mSUGRA (blue solid lines)~\protect\cite{MC3,MC4,MC5}.}
  \label{fig:predictions}
\end{figure}

As seen in the lower right panel of
Fig.~\ref{fig:predictions}, the cross section for spin-independent dark matter scattering on a proton may
be $\sim 10^{-45}$~cm$^2$~\cite{MC5}, 
within an order of magnitude of the present experimental limit, and within reach of
experiments now running or in preparation. These experiments may provide the keenest competition for the 
LHC in the search for supersymmetric particles. Note, however, that the two classes of experiment are quite
complementary. The LHC experiments may be able to discover missing-energy events and show that they 
are due to the production and decay of sparticles, but they will not be able to prove that the particles
carrying away the missing energy are completely stable and constitute the dark matter. On the other
hand, direct dark matter searches would be unable to prove that any detected dark matter particle
was supersymmetric. Only the combination of the two classes of experiment would be able to
establish a complete picture of SUSY, and the same is true in other scenarios for dark matter.

\section{The LHC Roulette Wheel}

The LHC is unique in my experience in that it is opening up the exploration of a new energy range up to a few TeV
where there are good reasons to expect new physics associated with the origin of particle masses and dark 
matter, but we do not know what form this new physics may take: Higgs, SUSY or something else. One can
compare the LHC start-up to a game of roulette: the wheel is now turning, the theoretical {\it `jeux sont 
faits'}, and it just remains to see where the ball will stop. The LHC has already told us about a few 
places where the ball does not stop, as described in the following paragraphs.

\begin{figure}[t!]
  \includegraphics[width=.5\textwidth]{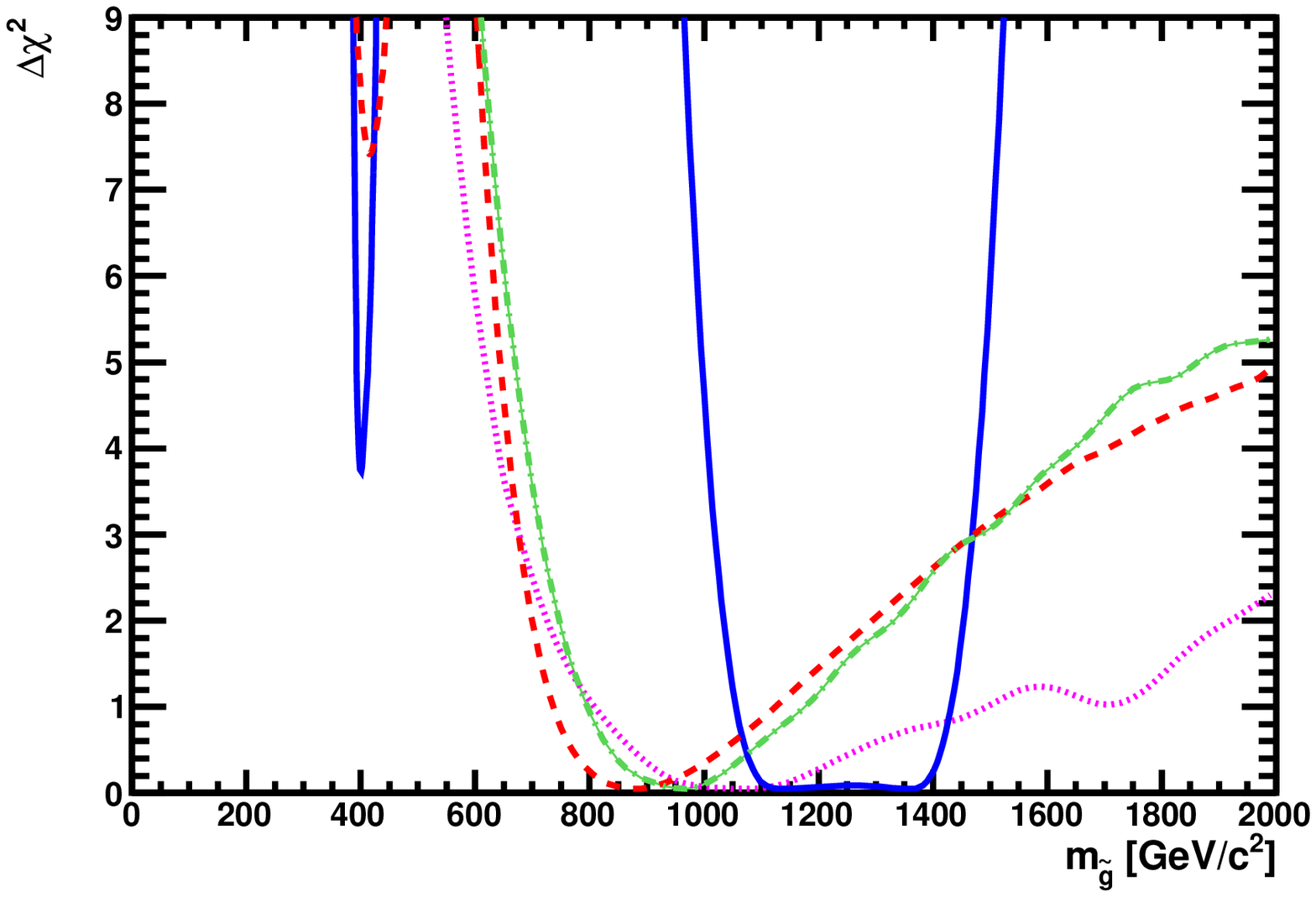}
  \includegraphics[width=.5\textwidth]{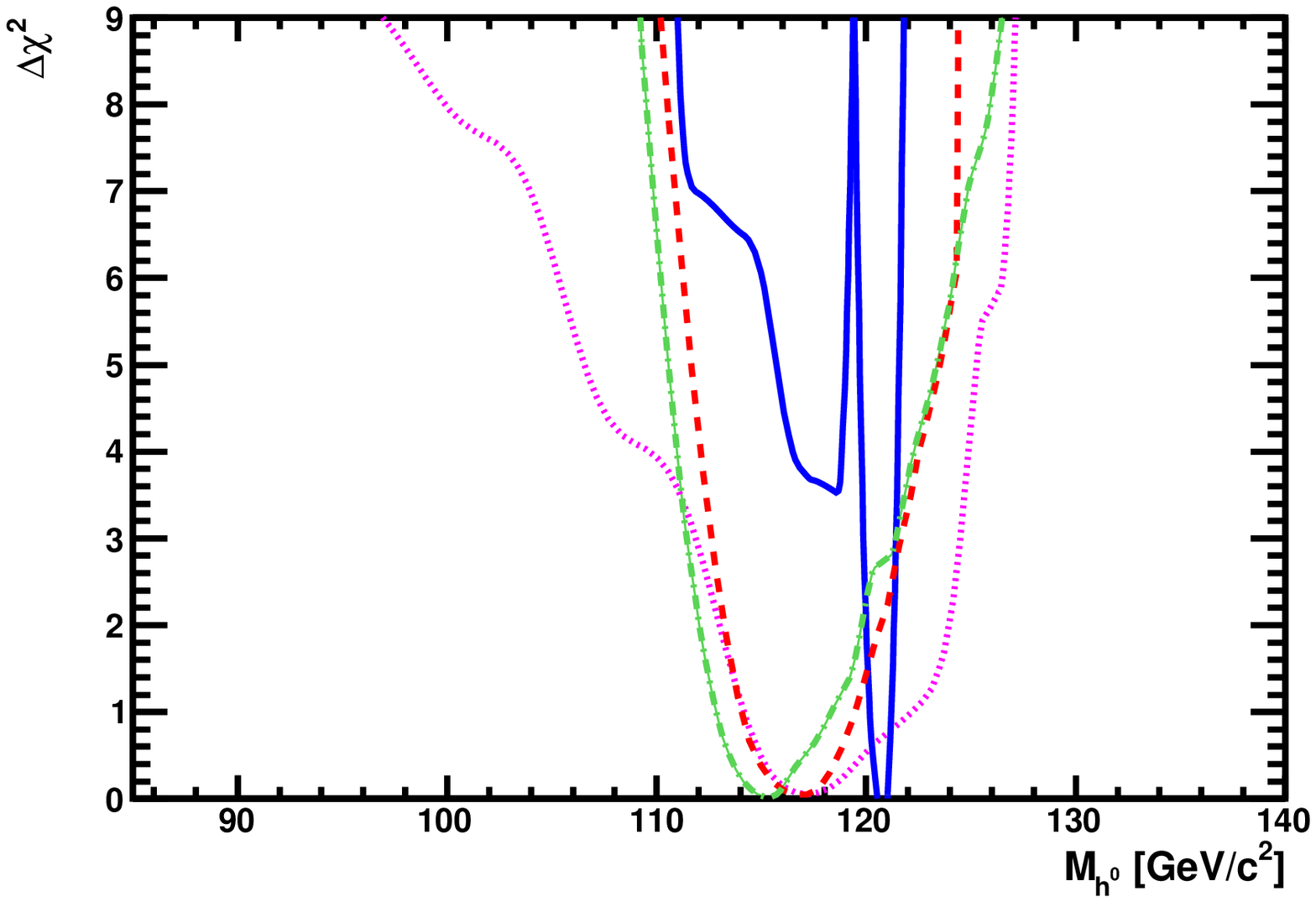}
  \end{figure}

\subsection{Composite quarks?}

One of the first LHC results that set limits on physics beyond the Standard Model that are stronger than those set by
previous experiments came from a search for excited quarks $q^*$ that might have been manufactured in $q + g$
collisions and decay {\it via} $q^* \to q + g$~\cite{ATLASq,CMSq}. 
As seen in the left panel of Fig.~\ref{fig:quark*}, these have now been
excluded with masses up to 1.58~TeV, much stronger than the limit of 0.87~TeV set at the Tevatron collider.

\subsection{String excitations?}

\begin{figure}
  \includegraphics[height=.29\textheight]{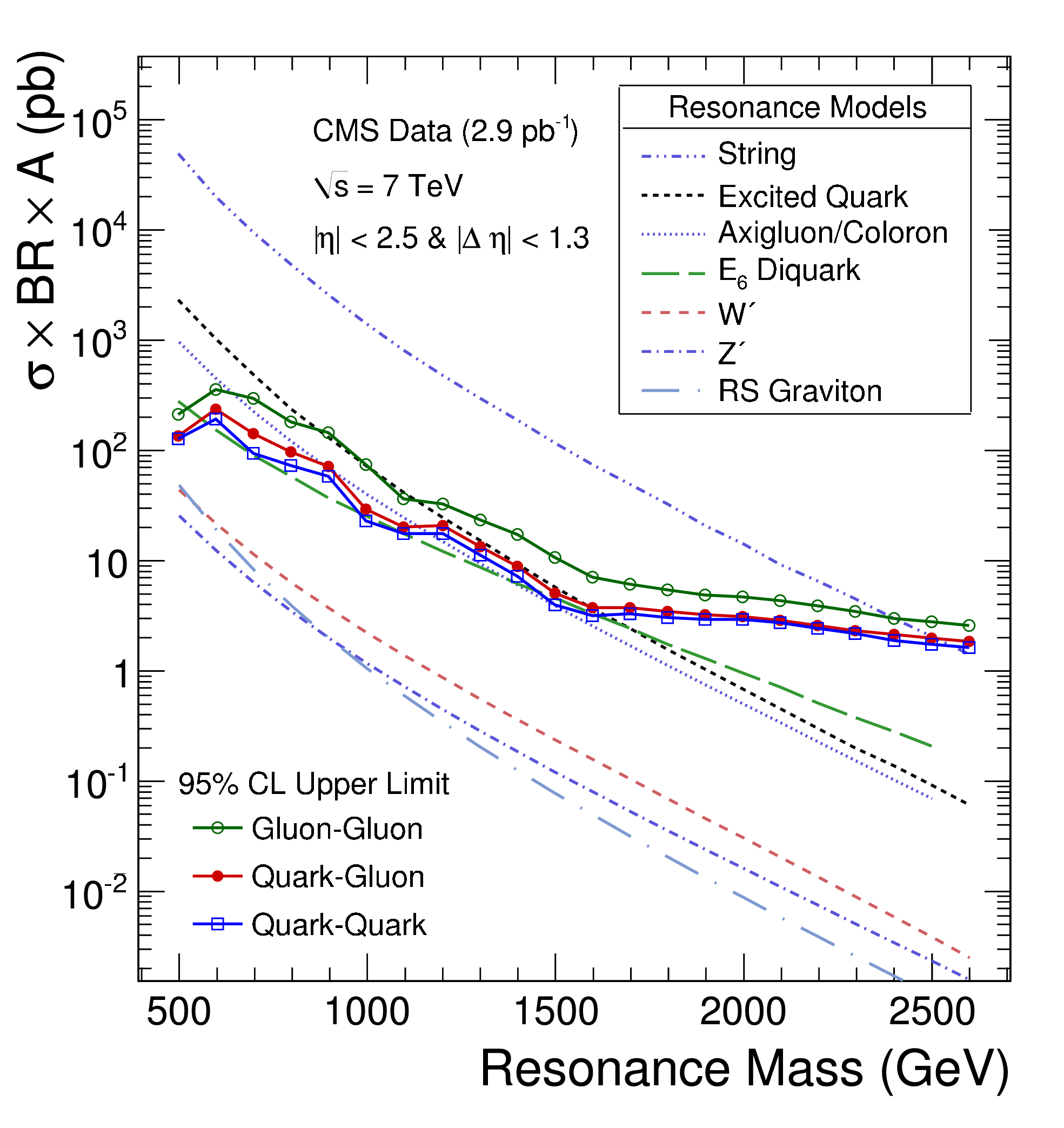}
  \includegraphics[height=.28\textheight]{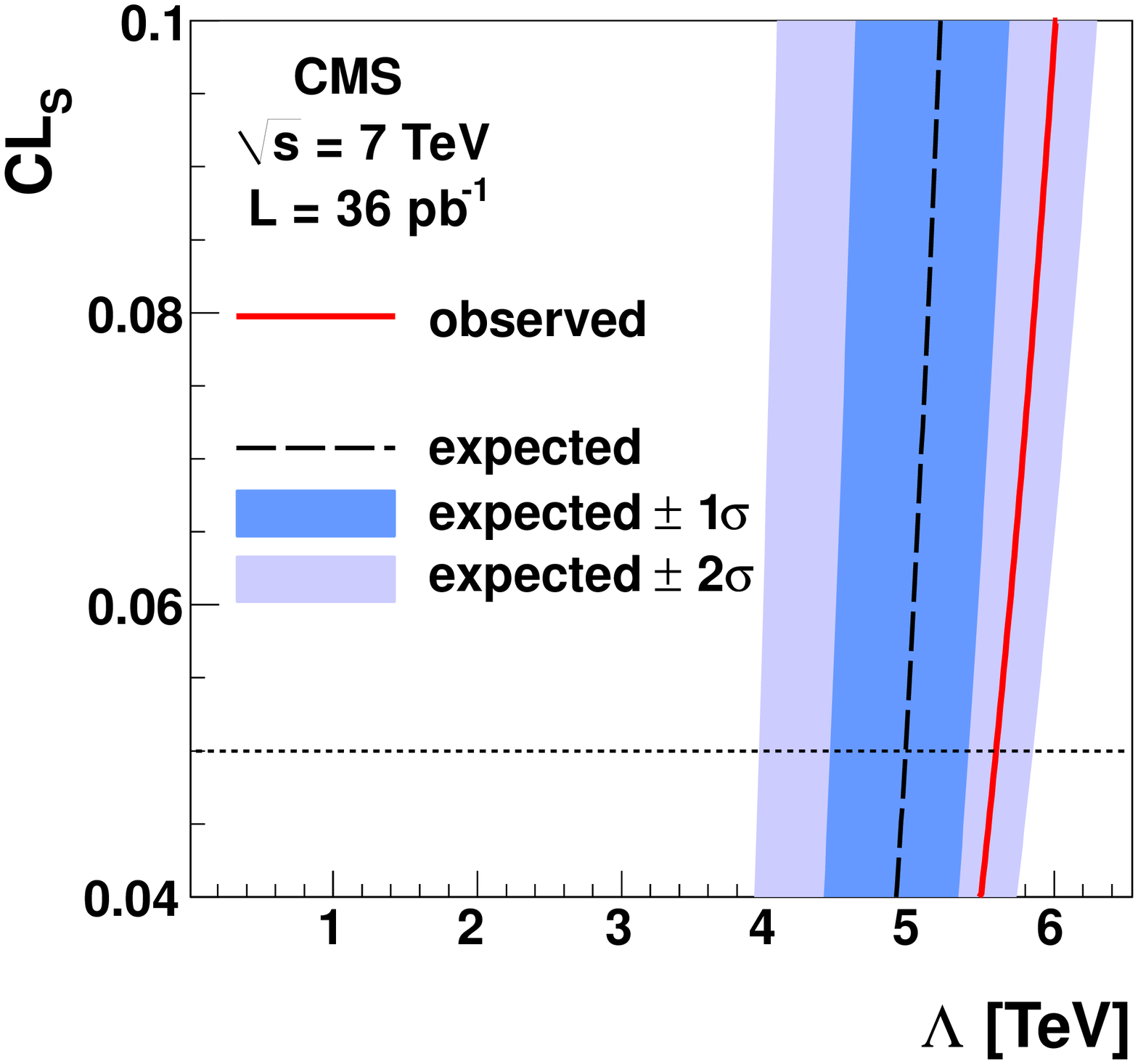}
  \caption{Left: upper limits on resonances in $g + g, q + g$ and $q + q$ scattering compared with the
  cross sections calculated in various scenarios, note in particular the limits on excited quarks and string
  resonances~\protect\cite{CMSq} (see also~\protect\cite{ATLASq}). 
  Right: limit on new contact interactions scaled by $\Lambda$, as obtained from a study of dijet
  angular distributions~\protect\cite{CMSangle} (see also~\protect\cite{CMSLambda,ATLASLambda}).}
  \label{fig:quark*}
\end{figure}

In some string scenarios, the scattering of quarks and gluons in the channels $q + q, q + g$ and $g + g$ may reveal
resonances at indistinguishable masses. The same LHC results shown in the left panel of Fig.~\ref{fig:quark*} also exclude this possibility up to a mass of 2.5~TeV~\cite{CMSq}, a limit that is also much stronger than previous constraints.

\subsection{Contact interactions?}

Another possibility in composite models is that there may be new, non-renormalizable contact interactions of the
form ${\bar q}q{\bar q}q$ and the like. These could show up {\it via} either deviations from the dijet invariant 
mass distributions calculated in QCD, or deviations from the expected angular distributions. The latter has
also been used to set limits stronger than in previous experiments, as seen in the right panel of 
Fig.~\ref{fig:quark*}~\cite{CMSangle} (see also~\cite{CMSLambda,ATLASLambda,EKS}).

\begin{figure}[b!]
  \includegraphics[height=.3\textheight]{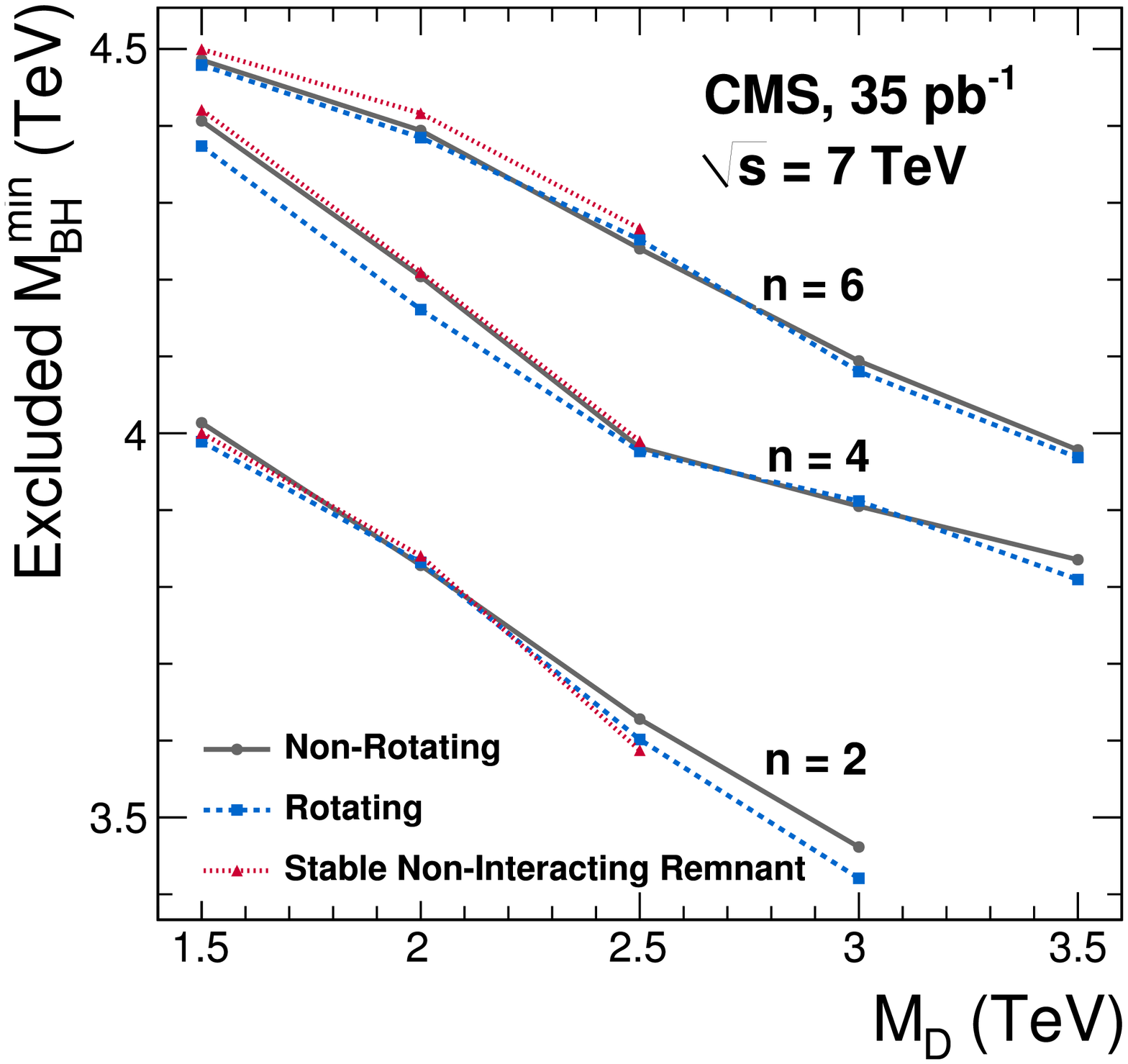}
  \caption{Excluded ranges of microscopic black hole masses under various assumptions about the number
  of extra dimensions $n$, the extra-dimensional Planck mass $M_D$, the angular momentum of the black
  hole and whether its decay leaves behind a (metastable) remnant~\protect\cite{CMSBH}.}
  \label{fig:LHCBH}
\end{figure}

\subsection{Microscopic black holes?}

In some theories with large extra dimensions, gravity may become strong at the TeV scale, in which case the 
high-energy collisions of quarks and gluons might produce microscopic black holes~\cite{LHCBH}. 
The theories that
predict such a possibility also predict that these microscopic black holes would decay very rapidly
through Hawking radiation. (This has not averted some unfounded speculations that LHC 
collisions might produce stable black holes capable of eating up the Earth, speculations
that are excluded by simple considerations of high-energy cosmic ray collisions on the
Earth and elsewhere in the Universe~\cite{LSAG}.) The production and decay of microscopic black
holes at the LHC has now been excluded over a large range of masses, as seen in
Fig.~\ref{fig:LHCBH}~\cite{CMSBH}.

\subsection{How else to probe string theory?}

A remarkable recent theoretical development has been the realization that the AdS/CFT correspondence
suggested by string theory could be used to calculate in simplified theories
properties of the quark-gluon matter produced in
relativistic heavy-ion collisions, starting with its viscosity~\cite{viscosity}.
Measurements of the viscosity of the medium produced in such collisions at RHIC have indicated that
it is remarkably low~\cite{lowvis}, far lower than that of the superfluid Helium cooling the LHC magnets,
and within a factor $\sim 3$ of the AdS/CFT lower limit. Early data from heavy-ion collisions seem to
confirm the low viscosity of the quark-gluon medium~\cite{ALICEv2}, and also to provide remarkable
evidence for large parton energy loss~\cite{ATLASquench,ALICEloss}. Is it too much to hope for some
quantitative tests of string ideas in heavy-ion collisions at the LHC?

\section{A Conversation with Mrs. Thatcher}

In 1982, just after the CERN ${\bar p} p$ collider started up, Mrs. Thatcher, 
the British Prime Minister at the time, came to visit CERN,
and I was introduced to her as a theoretical physicist. ``What exactly do you do?",
she asked in her inimitably intimidating manner. ``I think of things for experimentalists to look
for, and then I hope they find something different", I responded. Somewhat
predictably, Mrs. Thatcher asked ``Wouldn't it be better if they found what
you predicted?" My response was that "If they found exactly what the theorists
predicted, we would not be learning so much". As it happened, the
CERN ${\bar p} p$ collider found the $W^\pm$ and $Z^0$ particles, as expected.
Nevertheless, in much the same spirit as in 1982, I hope
(and indeed expect) that the LHC will become most famous for discovering
something NOT discussed in this talk!


\begin{theacknowledgments}
It is a pleasure to thank fellow members of the {\tt MasterCode} Collaboration for sharing the fun, and
the organizers for their kind invitation to speak at this interesting meeting.
\end{theacknowledgments}



\begin{thebibliography}{9}

\bibitem{Medellin}
M.~Bustamante, L.~Cieri and J.~Ellis,
{\it Beyond the Standard Model for Montaneros,}
  arXiv:0911.4409 [hep-ph].

\bibitem{LEPEWWG}
ALEPH, CDF, D0, DELPHI, L3, OPAL and SLD Collaborations, LEP and Tevatron Electroweak Working Groups, SLD Electroweak and Heavy Flavour Groups,
arXiv:0911.2604.

\bibitem{TDRs}
G.~Aad {\it et al.}  [The ATLAS Collaboration],
  ``Expected Performance of the ATLAS Experiment - Detector, Trigger and
  Physics,''
  arXiv:0901.0512;
G.~L.~Bayatian {\it et al.}, CMS Collaboration,  
{\it CMS Technical Design Report, Volume II: Physics Performance}, 
CERN-LHCC-2006-021, CMS-TDR-008-2 
J.~Phys.~G34, 995 (2007); see: {\tt http://cmsdoc.cern.ch/cms/cpt/tdr/}~.

\bibitem{LHCb}
B.~Adeva {\it et al.} [The LHCb Collaboration],
{\it Roadmap for selected key measurements of LHCb},
  arXiv:0912.4179 [hep-ex].
  
\bibitem{ALICE}
K.~Aamodt {\it et al.}, ALICE Collaboration, {\it The ALICE experiment at the CERN LHC},
J.~Inst. {\bf 3} (2008) S08002.

\bibitem{EB}
F.~Englert and R.~Brout,
  Phys.\ Rev.\ Lett.\  {\bf 13} (1964) 321.
  
\bibitem{Higgs}
P.~W.~Higgs,
  Phys.\ Lett.\  {\bf 12} (1964) 132 and
  Phys.\ Rev.\ Lett.\  {\bf 13} (1964) 508.
    
\bibitem{others}
See also P.~W.~Anderson,
  Phys.\ Rev.\  {\bf 130} (1963) 439 and
  G.~S.~Guralnik, C.~R.~Hagen and T.~W.~B.~Kibble,
  Phys.\ Rev.\ Lett.\  {\bf 13} (1964) 585.
For a recent review of the literature, see
See also L.~{\'A}lvarez-Gaum{\'e} and J.~Ellis, Nature Physics {\bf 7} (2011) 2.

\bibitem{LEPH}
S.~Schael {\it et al.}  [ALEPH, DELPHI, L3, OPAL
                       Collaborations and LEP Working Group for Higgs
                       boson searches],
  Eur.\ Phys.\ J.\  C {\bf 47} (2006) 547
  [arXiv:hep-ex/0602042].

\bibitem{TeVH}
CDF and D0 Collaborations,
  arXiv:1007.4587 [hep-ex].

\bibitem{GFitter}
GFitter Collaboration, {\tt http://gfitter.desy.de/}.

\bibitem{Dark}
See, for example: N.~A.~Bahcall, J.~P.~Ostriker, S.~Perlmutter and P.~J.~Steinhardt,
  Science {\bf 284} (1999) 1481
  [arXiv:astro-ph/9906463].
  
\bibitem{EHNOS}
J. Ellis, J.S. Hagelin, D.V. Nanopoulos, K.A. Olive
and M. Srednicki, Nucl. Phys. B {\bf 238} (1984) 453; see also
H. Goldberg, Phys. Rev. Lett. {\bf 50} (1983) 1419.

\bibitem{SUSY}
See, for example: P.~Fayet and S.~Ferrara,
  Phys.\ Rept.\  {\bf 32} (1977) 249.
  
\bibitem{hierarchy}
L.~Maiani, {\it All You Need To Know About The Higgs Boson}, Proceedings
of the Gif-sur-Yvette Summer School On Particle Physics, 1979, pp.1-52; G.~'t~Hooft, in {\it
Recent developments in Gauge Theories}, Proceedings of the NATO Advanced Study
Institute, Carg{\`e}se, 1979, eds. G.~'t~Hooft et al. (Plenum Press, NY, 1980);
E.~Witten,
  Phys.\ Lett.\ B {\bf 105} (1981) 267.

\bibitem{GUTs}
J. Ellis, S. Kelley and D.V. Nanopoulos, Phys. Lett. {\bf 260}
(1991) 131;
U. Amaldi, W. de Boer and H. Furstenau, Phys. Lett. {\bf B260} (1991) 447;
P. Langacker and M. Luo, Phys. Rev. {\bf D44} (1991) 817;
C.~Giunti, C.~W.~Kim and U.~W.~Lee,
  Mod.\ Phys.\ Lett.\  A {\bf 6} (1991) 1745.

\bibitem{SUSYH}
J.~R.~Ellis, G.~Ridolfi and F.~Zwirner,
  Phys.\ Lett.\ B {\bf 257} (1991) 83;
  Phys.\ Lett.\ B {\bf 262} (1991) 477;
  Yasuhiro Okada, Masahiro Yamaguchi and Tsutomu Yanagida,
{\em Phys. Lett.} B262, 54, 1991;
{\em Prog. Theor. Phys.} 85, 1, 1991;
  A.~Yamada,
  Phys.\ Lett.\ B {\bf 263}, 233 (1991);
  Howard~E. Haber and Ralf Hempfling,
{\em Phys. Rev. Lett.} 66, 1815, 1991;
M.~Drees and M.~M.~Nojiri,
  Phys.\ Rev.\ D {\bf 45} (1992) 2482;
  P.~H.~Chankowski, S.~Pokorski and J.~Rosiek,
  Phys.\ Lett.\ B {\bf 274} (1992) 191;
  Phys.\ Lett.\ B {\bf 286} (1992) 307;

\bibitem{noren}
J.~Iliopoulos and B.~Zumino,
  Nucl.\ Phys.\  B {\bf 76} (1974) 310;
  S.~Ferrara, J.~Iliopoulos and B.~Zumino,
  Nucl.\ Phys.\  B {\bf 77} (1974) 413.

\bibitem{Zerwas}
See, for example: G.~A.~Blair, W.~Porod and P.~M.~Zerwas,
  Eur.\ Phys.\ J.\  C {\bf 27} (2003) 263
  [arXiv:hep-ph/0210058].
  
\bibitem{Li}
See, for example, B.~Fields and S.~Sarkar, Particle Data Group, \\
{\tt http://pdg.lbl.gov/2010/reviews/rpp2010-rev-bbang-nucleosynthesis.pdf}.

\bibitem{Sakharov}
A.~D.~Sakharov, Pisma Zh. Eksp. Teor. Fiz. {\bf 5} (1967) 32 [JETP Lett. {\bf 5} (1967) 24].

\bibitem{CPV}
D.~Asner {\it et al.}, Heavy Flavour Averaging Group, arXiv:1010.1589 [hep-ex].

\bibitem{SUSYBG}
See M.~Carena, G.~Nardini, M.~Quiros and C.~E.~M.~Wagner,
  Nucl.\ Phys.\  B {\bf 812} (2009) 243
  [arXiv:0809.3760 [hep-ph]] and references therein.

\bibitem{LHCBH}
S.~B.~Giddings and S.~D.~Thomas,
  Phys.\ Rev.\  D {\bf 65} (2002) 056010
  [arXiv:hep-ph/0106219];
S.~Dimopoulos and G.~L.~Landsberg,
  Phys.\ Rev.\ Lett.\  {\bf 87} (2001) 161602
  [arXiv:hep-ph/0106295].

\bibitem{Webber}
C.~M.~Harris, M.~J.~Palmer, M.~A.~Parker, P.~Richardson, A.~Sabetfakhri and B.~R.~Webber,
  JHEP {\bf 0505} (2005) 053
  [arXiv:hep-ph/0411022].

\bibitem{Ketevi}
G.~Aad {\it et al.}, ATLAS Collaboration,
ATLAS-CONF-2011-005.

\bibitem{ATLASH}
E.~Nurse, for the ATLAS Collaboration, LHC End-Of-Year Jamboree, December 17th 2010,
{\tt http://indico.cern.ch/getFile.py/access?resId=0{\&}materialId=slides{\&}}
{\tt contribId=29{\&}sessionId=2{\&}subContId=5{\&}confId=114484}.

\bibitem{CMSH}
P.~Schieferdecker, for the CMS Collaboration, LHC End-Of-Year Jamboree, December 17th 2010,
{\tt http://indico.cern.ch/getFile.py/access?resId=0{\&}materialId=slides{\&}}
{\tt contribId=29{\&}sessionId=2{\&}subContId=4{\&}confId=114484}.

\bibitem{B2S}
Letter from R. Brinkman to M. Schochet, Jan. 6th, 2011, \\
{\tt http://www.fnal.gov/pub/today/Tevatron-brinkman-to-shochet.pdf}.

\bibitem{EEGHR}
See, for example: J.~Ellis, J.~R.~Espinosa, G.~F.~Giudice, A.~Hoecker and A.~Riotto,
  Phys.\ Lett.\  B {\bf 679} (2009) 369
  [arXiv:0906.0954 [hep-ph]].

\bibitem{Hinflation}
See, for example: F.~Bezrukov, A.~Magnin, M.~Shaposhnikov and S.~Sibiryakov,
  arXiv:1008.5157 [hep-ph].

\bibitem{EN}
J.~R.~Ellis and D.~V.~Nanopoulos,
  Phys.\ Lett.\  B {\bf 110} (1982) 44.
  
\bibitem{BG}
R.~Barbieri and R.~Gatto,
  Phys.\ Lett.\  B {\bf 110} (1982) 211.

\bibitem{Fayet}
P.~Fayet,
  Phys.\ Lett.\  B {\bf 69} (1977) 489.

\bibitem{LKP}
G.~Servant and T.~M.~P.~Tait,
  Nucl.\ Phys.\  B {\bf 650} (2003) 391
  [arXiv:hep-ph/0206071].
  
\bibitem{LTP}
H. C. Cheng and I. Low, {\bf JHEP} 0309 (2003) 051 and {\bf JHEP} 0408 (2004) 061.

\bibitem{EOSS}
Updated from  J.~R.~Ellis, K.~A.~Olive, Y.~Santoso and V.~C.~Spanos,
  Phys.\ Lett.\  B {\bf 565}, 176 (2003)
  [arXiv:hep-ph/0303043].

\bibitem{EMO}
J.~Ellis, A.~Mustafayev and K.~A.~Olive,
  Eur.\ Phys.\ J.\  C {\bf 69} (2010) 201
  [arXiv:1003.3677 [hep-ph]].
  
\bibitem{LEPsusy}
Joint Supersymmetry Working Group of the ALEPH, DELPHI, L3 and OPAL experiments,
{\tt http://lepsusy.web.cern.ch/lepsusy/}.

\bibitem{WMAP}
E.~Komatsu {\it et al.},
  arXiv:1001.4538 [astro-ph.CO] and references therein.

\bibitem{g-2}
G.~W.~Bennett {\it et al.}  [Muon g-2 Collaboration],
  Phys.\ Rev.\ Lett.\  {\bf 92} (2004) 161802
  [arXiv:hep-ex/0401008].

\bibitem{Davier}
See M.~Davier, A.~Hoecker, B.~Malaescu and Z.~Zhang,
  arXiv:1010.4180 [hep-ph] and references therein.

\bibitem{MC4}
O.~Buchmueller {\it et al.},
  arXiv:1011.6118 [hep-ph].
  
\bibitem{CMSlimit}
V.~Khachatryan {\it et al.}, CMS Collaboration, arXiv:1101.1628 [hep-ex].

\bibitem{ATLASlimit}
G.~Aad {\it et al.} ATLAS Collaboration,
arXiv:1102.2357 [hep-ex].

\bibitem{MC5}
O.~Buchmueller {\it et al.},
  arXiv:1102.4585 [hep-ph].
  
\bibitem{ATLASsusy}
ATLAS Collaboration, {\tt http://cdsweb.cern.ch/record/1278474/} \\
{\tt files/ATL-PHYS-PUB-2010-010.pdf}; see also \\
CMS Collaboration, CMS-NOTE-2010-008, {\tt https://twiki.cern.ch/} \\
{\tt twiki/bin/view/CMSPublic/PhysicsResultsSUS}.

\bibitem{CMSsusy}
CMS Collaboration, CMS NOTE 2010/008 available from \\
{\tt https://twiki.cern.ch/twiki/bin/view/CMSPublic/PhysicsResultsSUS}.

\bibitem{MC3}
O.~Buchmueller {\it et al.},
  JHEP {\bf 0809} (2008) 117
  [arXiv:0808.4128 [hep-ph]];
  Eur.\ Phys.\ J.\  C {\bf 64} (2009) 391
  [arXiv:0907.5568 [hep-ph]].

\bibitem{ATLASq}
G.~Aad {\it et al.}  [ATLAS Collaboration],
  Phys.\ Rev.\ Lett.\  {\bf 105} (2010) 161801
  [arXiv:1008.2461 [hep-ex]].

\bibitem{CMSq}
V.~Khachatryan {\it et al.}  [CMS Collaboration],
  Phys.\ Rev.\ Lett.\  {\bf 105} (2010) 211801
  [arXiv:1010.0203 [hep-ex]].
  
\bibitem{CMSangle}
  V.~Khachatryan {\it et al.}  [CMS Collaboration],
  arXiv:1102.2020 [hep-ex].
  
\bibitem{CMSLambda}
V.~Khachatryan {\it et al.}  [CMS Collaboration],
  Phys.\ Rev.\ Lett.\  {\bf 105} (2010) 262001
  [arXiv:1010.4439 [hep-ex]].
  
\bibitem{ATLASLambda}
G.~Aad {\it et al.}  [ATLAS Collaboration],
  Phys.\ Lett.\  B {\bf 694} (2011) 327
  [arXiv:1009.5069 [hep-ex]].
  
\bibitem{EKS}
Hadronic antenna patterns due to colour radiation may be a useful supplementary tool to identify unusual
mechanisms for large-$E_T$ jet production: see
J.~R.~Ellis, V.~A.~Khoze and W.~J.~Stirling,
  Z.\ Phys.\  C {\bf 75} (1997) 287
  [arXiv:hep-ph/9608486].
  
\bibitem{LSAG}
J.~R.~Ellis, G.~Giudice, M.~L.~Mangano, I.~Tkachev and U.~Wiedemann,
  J.\ Phys.\ G {\bf 35} (2008) 115004
  [arXiv:0806.3414 [hep-ph]].
  
\bibitem{CMSBH}
V.~Khachatryan {\it et al.}  [CMS Collaboration],
  arXiv:1012.3375 [hep-ex].

\bibitem{viscosity}
P.~Kovtun, D.~T.~Son and A.~O.~Starinets,
  Phys.\ Rev.\ Lett.\  {\bf 94} (2005) 111601
  [arXiv:hep-th/0405231].
  
\bibitem{lowvis}
See, for example: H.~Song, S.~A.~Bass, U.~W.~Heinz, T.~Hirano and C.~Shen,
  arXiv:1011.2783 [nucl-th].

\bibitem{ALICEv2}
K.~Aamodt {\it et al.}  [The ALICE Collaboration],
  arXiv:1011.3914 [nucl-ex].

\bibitem{ATLASquench}
G.~Aad {\it et al.}  [Atlas Collaboration],
  arXiv:1011.6182 [hep-ex].

\bibitem{ALICEloss}
K.~Aamodt {\it et al.}  [ALICE Collaboration],
  Phys.\ Lett.\  B {\bf 696} (2011) 30
  [arXiv:1012.1004 [nucl-ex]].

\end{thebibliography}
\end{document}